\documentclass[a4paper,fleqn,usenatbib]{mnras}

\usepackage{newtxtext,newtxmath}

\usepackage[T1]{fontenc}
\usepackage{ae,aecompl}

\usepackage{graphicx}	
\usepackage{amsmath}	
\usepackage{amssymb}	
\usepackage{threeparttable}
\usepackage{booktabs}
\usepackage{verbatim}
\usepackage[all]{hypcap}
\usepackage[usenames, dvipsnames]{color}
\usepackage[utf8]{inputenc}

\title[Assembly bias in KiDS+GAMA]{A KiDS weak lensing analysis of assembly bias in GAMA galaxy groups}

\author[A. Dvornik et al.]{Andrej Dvornik,$^{1}$\thanks{E-mail: dvornik@strw.leidenuniv.nl}
Marcello Cacciato,$^{1}$
Konrad Kuijken,$^{1}$
Massimo Viola,$^{1}$ \newauthor
Henk Hoekstra,$^{1}$ 
Reiko Nakajima,$^{2}$
Edo van Uitert,$^{3}$ 
Margot Brouwer,$^{1}$
Ami Choi,$^{7}$ \newauthor
Thomas Erben,$^{2}$
Ian Fenech Conti,$^{4,5}$ 
Daniel J. Farrow,$^{6}$ 
Ricardo Herbonnet,$^{1}$ \newauthor
Catherine Heymans,$^{7}$ 
Hendrik Hildebrandt,$^{2}$  
Andrew. M. Hopkins,$^{11}$ 
John McFarland,$^{8}$ \newauthor
Peder Norberg,$^{10}$ 
Peter Schneider,$^{2}$
Crist{\'o}bal Sif{\'o}n,$^{9}$
Edwin Valentijn,$^{8}$ 
Lingyu Wang$^{8, 12}$
\\
$^{1}$Leiden Observatory, Leiden University, Niels Bohrweg 2, 2333 CA Leiden, The Netherlands.\\
$^{2}$Argelander-Institut f{\"u}r Astronomie, Auf dem H{\"u}gel 71, 53121 Bonn, Germany.\\
$^{3}$University College London, Gower Street, London WC1E 6BT, UK.\\
$^{4}$Institute of Space Sciences and Astronomy (ISSA), University of Malta, Msida, MSD 2080, Malta.\\
$^{5}$Department of Physics, University of Malta, Msida, MSD 2080, Malta.\\
$^{6}$Max-Planck-Institut f{\"u}r extraterrestrische Physik, Postfach 1312 Giessenbachstrasse, D-85741 Garching, Germany.\\
$^{7}$SUPA, Institute for Astronomy, University of Edinburgh, Royal Observatory, Blackford Hill, Edinburgh EH9 3HJ, UK.\\
$^{8}$Kapteyn Astronomical Institute, P.O. Box 800, 9700 AV Groningen, The Netherlands.\\
$^{9}$Department of Astrophysical Sciences, Peyton Hall, Princeton University, Princeton, NJ 08544, USA.\\
$^{10}$ICC \& CEA, Department of Physics, Durham University, South Road, Durham DH1 3LE, UK.\\
$^{11}$Australian Astronomical Observatory, P.O. Box 915, North Ryde, NSW 1670, Australia.\\
$^{12}$SRON Netherlands Institute for Space Research, Landleven 12, 9747 AD Groningen, The Netherlands.
}

\date{Accepted 2017 March 20. Received 2017 March 20; in original form 2016 December 09}

\pubyear{2016}

\begin{document}
\label{firstpage}
\pagerange{\pageref{firstpage}--\pageref{lastpage}}
\maketitle

\begin{abstract}
We investigate possible signatures of halo assembly bias for spectroscopically selected galaxy groups from the GAMA survey using weak lensing measurements from the spatially overlapping regions of the deeper, high-imaging-quality photometric KiDS survey. We use GAMA groups with an apparent richness larger than 4 to identify samples with comparable mean host halo masses but with a different radial distribution of satellite galaxies, which is a proxy for the formation time of the haloes. We measure the weak lensing signal for groups with a steeper than average and with a shallower than average satellite distribution and find no sign of halo assembly bias, with the bias ratio of $0.85^{+0.37}_{-0.25}$, which is consistent with the $\Lambda$CDM prediction. Our galaxy groups have typical masses of $10^{13} M_{\odot}/h$, naturally complementing previous studies of halo assembly bias on galaxy cluster scales.

\end{abstract}

\begin{keywords}
gravitational lensing: weak -- methods: statistical -- surveys -- galaxies: haloes -- dark matter -- large-scale structure of Universe.
\end{keywords}


\section{Introduction}
\label{sec:intro}

In the standard cold dark matter and cosmological constant dominated ($\Lambda$CDM) cosmological framework, structure formation in the Universe is mainly driven by the dynamics
of cold dark matter. The gravitational collapse of dark matter density fluctuations and their subsequent virialization leads to the formation of dark matter haloes
from the highest density peaks in the initial Gaussian random density field \citep[e.g.][and the references therein]{Mo2010}. As dark matter haloes trace the 
underlying mass distribution, the halo bias (the relationship between the spatial distribution of dark matter haloes and the underlying dark matter density field) is naively expected to depend only on the halo mass, and can be used to predict the large-scale clustering of the dark matter haloes \citep{Zentner2013, Hearin2015}.

However, cosmological N-body simulations have shown that the abundance and clustering of the haloes depend on properties other than the halo mass alone. 
These for instance include formation time and concentration \citep{Wechsler2006, Gao2007, Dalal2008, Wang2009, Lacerna2014}. The dependence of the spatial distribution of dark matter 
haloes on any of those properties, or on any property beside mass, it is commonly called \emph{halo assembly bias} \citep{Hearin2015}. 

Cosmological N-body simulations indicate that the origin of halo assembly bias is twofold. While for the high-mass haloes the assembly bias comes purely from 
the statistics of density peaks \citep[related to the curvature of Lagrangian peaks in the initial Gaussian random density field;][]{Dalal2008}, the origin of halo assembly bias for low-mass haloes is rather a signature of cessation of mass accretion onto haloes \citep{Wang2009, Zentner2013}. 

As galaxies are biased tracers of the underlying dark matter distribution, halo assembly bias, to some extent, violates the standard halo occupation models, 
which in most cases assume that the halo mass alone can completely describe 
the statistical properties of galaxies residing in such dark matter haloes at a given time \citep{Leauthaud2011, Bosch2012, Cacciato2013}, and are used to connect the galaxies with their parent haloes in which they are formed. The central quantity upon which halo occupation models are built, is the probability of a halo hosting a given number of galaxies, given its halo mass. Assembly bias will thus violate 
the mass-only assumption, and those models will introduce systematic errors when predicting the lensing signal and/or clustering measurements of galaxies, 
groups and clusters when split into subsamples of a different secondary observable (for instance, concentration) \citep{Zentner2013}. Because of that, there has been an increased effort in the 
last couple of years to accommodate models for assembly bias, by expanding them to allow for secondary properties to govern the occupational 
distributions \citep{Hearin2015}.

It has also been shown that assembly bias introduces a bimodality to the halo bias function -- the function relating the clustering of matter with the observed clustering of haloes (i.e. one gets two functions, whose properties differ by the secondary observable) -- but preserving the overall mass 
dependence \citep[the more massive the halo, the larger the split and thus the assembly bias;][]{Gao2007}. As halo assembly bias can be a signature of a multitude of secondary properties 
(formation time, concentration, host galaxy colour, amongst others), further study across multiple mass scales (from galaxies to galaxy clusters) using the same proxy is needed, as the mass dependence of halo assembly bias is not completely determined observationally. 

Several studies have presented observational evidence of halo assembly bias. \citet{Yang2006} showed that 
at fixed halo mass, galaxy clustering increases with decreasing star formation rate (SFR) and that the reshuffling of observational quantities (dynamical mass and the total stellar mass) affects 
the clustering signal by up to $10\%$. Their results are in agreement with the findings from \citet{Gao2005}, who used results from the Millennium simulation \citep{Springel2005}. Similar results were more 
recently obtained by \citet{Tinker2012} using observations of the COSMOS field. They find that the stellar mass of the star-forming 
galaxies, residing in galaxy groups, is a factor of 2 lower than for passive galaxies residing in halos with the same mass. Moreover, a similar trend is observed 
when they divide the population of galaxies by their morphology (for details see the definition therein), emphasising the significantly different clustering amplitudes of the two observed samples. 
On the other hand, \citet{Lin2015} investigated some of these claims on galaxy scales using SDSS DR7 data \citep{Abazajian2009} and found no evidence for halo assembly bias, concluding that the observed differences in 
clustering were due to contamination from satellite galaxies. 

More recently, \citet{Miyatake2015} used galaxy-galaxy lensing and clustering measurements of more than $8000$ SDSS galaxy clusters with typical halo masses 
of \mbox{\textasciitilde{} $2 \times 10^{14} M_{\odot}/h$}, found using 
the redMaPPer method \citep{Rykoff2014}. They divided the clusters into two subsamples according to the radial distribution of the photometrically selected satellite 
galaxies from the brightest cluster galaxy. They found that the halo bias of clusters of the same halo mass but with different spatial distributions of satellite galaxies, differs up to $2.5\sigma$ in 
weak lensing, and up to $4.6\sigma$ in clustering measurements. \citet{Zu2016} argue that the detection of halo assembly bias by \citet{Miyatake2015} is driven purely by projection effect, and they show that the effects is smaller and consistent with $\Lambda$CDM predictions.

We aim to investigate whether signatures of halo assembly bias are present in galaxy groups with typical masses of $10^{13} M_{\odot}/h$, using measurements of the weak gravitational lensing signal. Specifically we use \emph{spectroscopically} selected galaxy groups from the 
GAMA survey \citep{Driver2011} and measure the weak lensing signal from the spatially overlapping regions of the deeper, high imaging 
quality photometric KiDS survey \citep{Kuijken2015, DeJong2015}. 
As the GAMA survey provides us with spectroscopic information on the group membership, any potential projection effects are much more confined. In order to see if the two population of groups have the clustering properties consistent with what halo masses dictate, we need to 
know the halo masses of the two populations. Because of that we interpret 
the measured signal in the context of the halo model \citep{Seljak2000, Cooray2002, Bosch2012, Cacciato2013}.

The outline of this paper is as follows. In Section \ref{sec:lensing} we describe the basics of the weak lensing theory, and we describe the data 
and sample selection in Section \ref{sec:data}. The halo model is described in Section \ref{sec:halomodel_intro}. In Section \ref{sec:results} 
we present the galaxy-galaxy lensing results. 
We conclude and discuss in Section \ref{sec:conclusions}. Throughout the paper we use the following cosmological parameters entering in the calculation of the distances and in the halo model 
\citep{PlanckCollaboration2014}: $\Omega_{\text{m}} = 0.315$, $\Omega_{\Lambda} = 0.685$, $\sigma_{8} = 0.829$, $n_{\text{s}} = 0.9603$ and $\Omega_{\text{b}}h^{2} = 0.02205$. All the measurements presented in the paper are in comoving units.

\section{Weak galaxy-galaxy lensing theory}
\label{sec:lensing}

Matter inhomogeneities deflect light rays of distant objects along their path. This effect is called gravitational lensing. 
As a consequence the images of distant objects (sources) appear to be tangentially distorted around foreground galaxies (lenses). The strength of the distortion is proportional to the amount of mass associated with the lenses and it is stronger in the proximity of the centre of the overdensity and becomes weaker at larger transverse distances
\citep[for a thorough review, see][]{Bartelmann1999}. 

Under the assumption that source galaxies have 
an intrinsically random ellipticity, weak gravitational lensing then introduces a coherent tangential distortion. The typical change in ellipticity due to gravitational 
lensing is much smaller than the intrinsic ellipticity of the source, even in the case of clusters of galaxies, but this can be overcome by 
averaging the shapes of many background galaxies. 

Weak gravitational lensing from a galaxy halo of a single galaxy is too weak 
to be detected. One therefore relies on a statistical approach in which one stacks the contributions from different lens galaxies, selected by similar 
observational properties (e.g. stellar masses, luminosities or in our case, the properties of the host of the satellite galaxies). Average halo properties, such as 
halo masses and large-scale halo biases, are then inferred from the resulting high signal-to-noise ratio measurements. This technique is commonly referred to as galaxy-galaxy lensing, and it is used as a method to 
measure statistical properties of dark matter halos around galaxies.

Given its statistical nature, galaxy-galaxy lensing can be considered as a measurement of the cross-correlation of galaxies and the matter density field:
\begin{equation}
\label{eq:Corr}
\xi_{\rm g,m}(\vert \textbf{r} \vert)=\langle \delta_{\rm g}(\textbf{x})\delta_{\rm m}(\textbf{x}+\textbf{r}) \rangle_{\textbf{x}}\, ,
\end{equation}
where $\mathrm{\delta_{\rm g}}$ is the galaxy density contrast, $\mathrm{\delta_{\rm m}}$ the matter density contrast, $\textbf{r}$ is the three-dimensional comoving separation and $\textbf{x}$ the position of the galaxy. From Equation \ref{eq:Corr} one can obtain the projected 
surface mass density around galaxies which, in the distant observer approximation, takes the form of an Abel transform:
\begin{equation}
\label{eq:projectCorr}
\Sigma(R) = 2\overline{\rho}_{\text{m}} \int^{\infty}_{R} \xi_{\text{g,m}}(r) \frac{r\, \mathrm{d}r}{\sqrt{r^{2} - R^{2}}}\, ,
\end{equation}
where $R$ is the comoving projected separation from the galaxy,  $\mathrm{\bar{\rho}_{m}}$ 
is the mean comoving density of the Universe and $r$ is the 3D comoving separation.\footnote{Throughout the paper we assume that the averaged mass profile of haloes is spherically symmetric, since we measure the lensing signal from a stack of many different haloes with different orientations, which averages out any potential halo triaxiality.} 
Being sensitive to density contrasts, gravitational lensing is actually a measure of the excess surface mass density (ESD):
\begin{equation}
\label{eq:ESD}
\Delta \Sigma (R)= {\bar \Sigma} (\le R)- \Sigma (R) \, , 
\end{equation}
where ${\bar \Sigma} (\le R)$ follows from:
\begin{equation}
\label{eq:SinR}
{\bar \Sigma} (\le R) = \frac{2}{R^2} \int_0^R \Sigma (R') \, R'  \,  \mathrm{d}R' \, .
\end{equation}
The ESD can finally be related to the tangential shear $\gamma_{\rm t}$ of background objects, which is the main lensing observable:
\begin{equation}
\label{eq:deltaSigmaTheo}
\Delta \Sigma (R)=\gamma_{\rm t} (R) \Sigma_{\rm cr} \, ,
\end{equation}
with
\begin{equation}
\label{eq:SigmaCrit}
\Sigma_{\rm cr}=\frac{c^2}{4\pi G} \frac{D(z_{\rm s})}{D(z_{\rm l})D(z_{\rm l},z_{\rm s})} \, ,
\end{equation}
the critical surface mass density, a geometrical factor accounting for the lensing efficiency. In the above equation, $D(z_{\rm l})$ is the angular diameter distance 
to the lens, $D(z_{\rm l},z_{\rm s})$ the angular diameter distance between the lens and the source and $D(z_{\rm s})$ the angular diameter 
distance to the source. In this equation $c$ denotes the speed of light and $G$ the gravitational constant.
In this work, the distances are evaluated using spectroscopic redshifts for the lenses and photometric redshifts for the sources.

Predictions on ESD profiles can be obtained by using the halo model of structure formation \citep{Seljak2000, Peacock2000, Cooray2002, Bosch2012, Mead2015} and we will 
base the interpretation of the measurements on this framework, which is presented in Section \ref{sec:halomodel_intro}.

\section{Data and sample selection}
\label{sec:data}

\subsection{Lens galaxy selection}
\label{sec:lenses}

The foreground galaxies used in this lensing analysis are taken from the Galaxy And Mass Assembly (hereafter GAMA) survey \citep{Driver2011}. 
GAMA is a spectroscopic survey carried out on the Anglo-Australian Telescope with the AAOmega spectrograph. Specifically, we use the information of GAMA galaxies from three equatorial regions, G9, G12 and G15 
from the GAMA II data release \citep{Liske2015}. We do not use the G02 and G23 regions, due to the fact that the first one does not overlap with KiDS and the second one uses a different target selection compared to the one used in the equatorial regions. These equatorial regions encompass \mbox{\textasciitilde{} 180 deg$^2$}, 
containing $180\,960$ galaxies (with $nQ > 3$, where the $nQ$ is a measure of redshift quality) and are highly complete down to a Petrosian $r$-band magnitude $r = 19.8$. 
For weak lensing measurements, we can use all the galaxies in the three equatorial regions as potential lenses.

We use the GAMA galaxy group catalogue version 7 \citep{Robotham2011} to separate galaxies into centrals and satellites. The centrals are used as centre of the haloes in the lensing analysis, while the distribution of satellites is used to separate haloes with an early and late formation time. The group catalogue is constructed with a Friends-of-Friends (FoF) algorithm that takes into account the projected and line-of-sight separations, and has been carefully calibrated against mock catalogues \citep{Robotham2011}, which were produced using the Millennium simulation \citep{Springel2005}, 
populated with galaxies according to the semi-analytical model by \citet{Bower2006}.

We select central galaxies residing in 
GAMA groups (the definition of the central galaxy used in this paper is the Brightest Cluster Galaxy\footnote{As shown in \citet{Robotham2011}, 
the iterative centre is the most accurate tracer of the centre of group, but using BCG as a tracer is not very different from it.} -- BCG) to trace the centres of the groups. We select all groups with an apparent richness\footnote{$N_{\text{Fof}}$ is defined by the number of GAMA galaxies associated with the group and it is dependent on the group selection function.} ($N_{\text{FoF}}$) larger than $N_{\text{FoF}} = 4$, 
covering a redshift range $0.03 \le z < 0.33$. With this apparent richness cut we minimise 
the fraction of spurious groups and the redshift cut provides a more reliable group sample (above the redshift of $z \sim 0.3$, the linking length used in the FoF algorithm can become excessively large). This selection yields $2061$ galaxy groups. If we include all the GAMA groups up to the redshift of $z = 0.5$, the final results do not change significantly, apart from having a higher signal-to-noise ratio in the lensing measurements, a result of having \mbox{\textasciitilde{} $200$} more galaxies in that sample. We thus opt for a cleaner sample of galaxy groups, whose membership is better under control.

As a proxy for the halo assembly bias signatures of our galaxy groups we employ the average projected 
separation of satellite galaxies, $\langle R \rangle$, from the central. The radial distribution of satellite galaxies is connected to the halo concentration and thus with the halo formation time, as shown in simulations \citep{Duffy2011, Bhattacharya2011}. This measurement is naturally given by the FoF algorithm run on the GAMA survey. 

\begin{figure*}
	\centering
	\includegraphics[width=17cm]{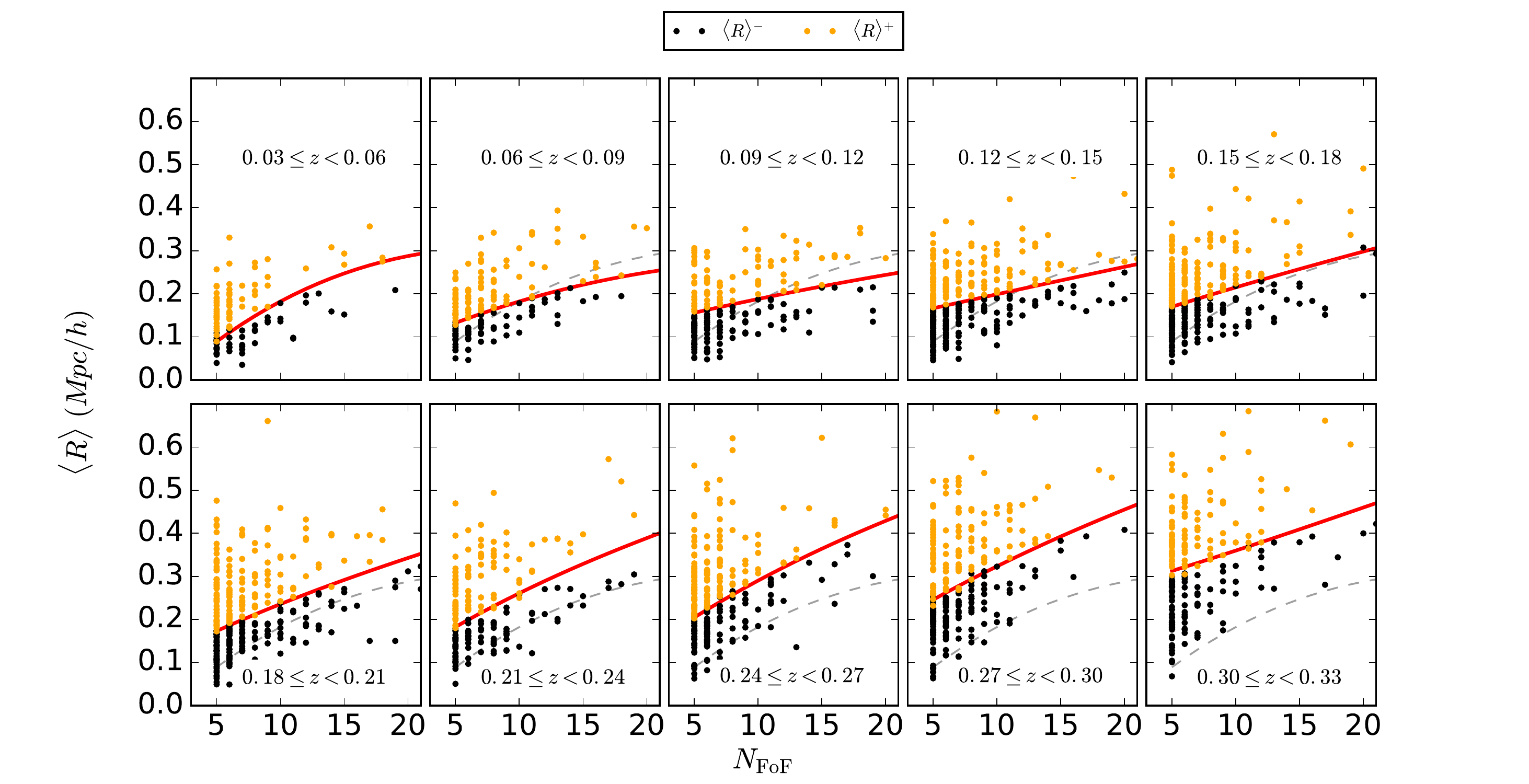}
 	\caption{Selection of GAMA groups with apparent richness $N_{\text{FoF}} \ge 5$ and redshift $0.03 \le z < 0.33$. In each panel groups are further split by the average projected distance, $\langle R \rangle$, of their satellite galaxies using a spline fit for the median of $\langle R \rangle$ (red curves). For brevity, we show only the apparent richnesses up to $20$. We plot the spline fit from the first redshift bin in all other bins in grey dashed lines. They are used to guide one's eye to see how spline changes from bin to bin.}
	\label{fig:selection}
\end{figure*}

Furthermore, we use this proxy to split our 
sample of central galaxies into two. We take 10 equally linearly spaced bins in $z$ and 15 in $N_{\text{FoF}}$ and 
perform a cubic spline fit for the median $\langle R \rangle$ as a function of $z$ and $N_{\text{FoF}}$ (see Figure \ref{fig:selection}). 

\begin{figure*}
	\centering
	\includegraphics[width=\textwidth]{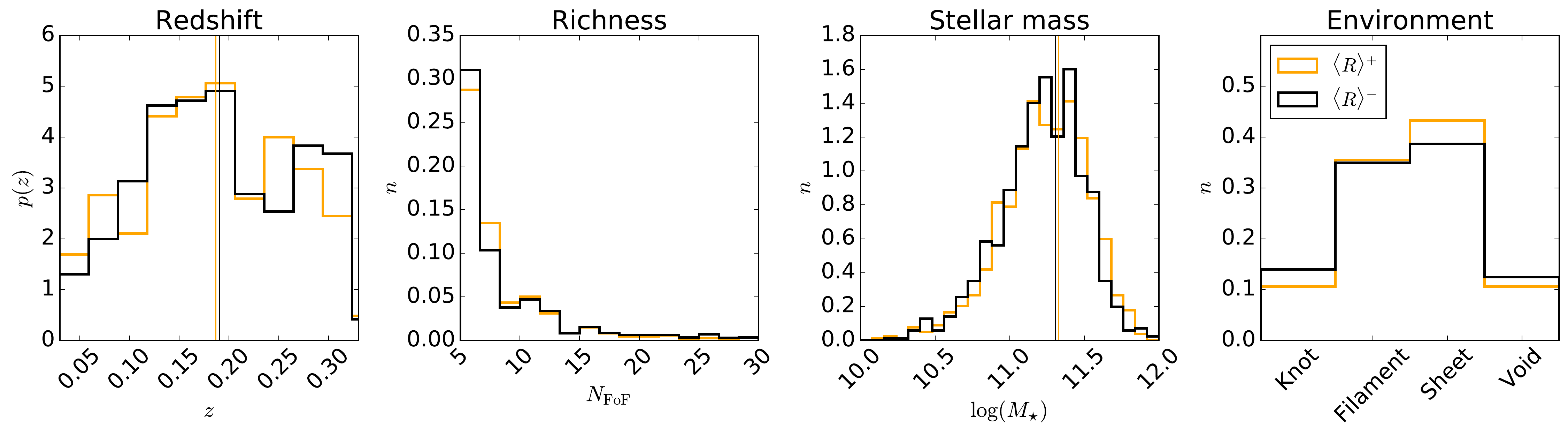}
 	\caption{\emph{Left panel:} Redshift distributions of the GAMA groups used in this paper for both the $\langle R \rangle^{+}$ and the $\langle R \rangle^{-}$ samples, shown as orange and black histograms. \emph{Middle left panel:} Apparent richness distributions of the GAMA groups used in this paper for both the $\langle R \rangle^{+}$ and the $\langle R \rangle^{-}$ samples. \emph{Middle right panel:} Stellar mass distributions of the GAMA groups used in this paper for both the $\langle R \rangle^{+}$ and the $\langle R \rangle^{-}$ samples. \emph{Right panel:} Distribution of the galaxy groups in different cosmic environments. The solid orange and black vertical lines indicate the median of the redshift and stellar mass distributions for the $\langle R \rangle^{+}$ and $\langle R \rangle^{-}$ sample, respectively.}
	\label{fig:distributions}
\end{figure*}

The spline fit gives us a limit between the central galaxies with satellites that are on average further apart from 
(upper half -- hereafter $\langle R \rangle^{+}$), 
or closer to (lower half -- hereafter $\langle R \rangle^{-}$) the BCG. 
The $\langle R \rangle^{+}$ sample has $987$ galaxy groups and the $\langle R \rangle^{-}$ sample $1074$ galaxy groups. 
This provides us, by construction, with two samples that have similar redshift, richness and stellar mass distributions, as can be seen in Figure \ref{fig:distributions}. 
The median stellar masses and redshifts are listed in Table \ref{tab:sample_properties}. As the dark matter haloes are located in different cosmic environments, we also want to check for the presence of apparent trends in our two samples with their environments.

\citet{Brouwer2016} presented a study of galaxies residing in different cosmic environments and they find a clear correlation of the halo bias with the cosmic environment of the haloes the galaxies are residing in. We check for the presence of apparent trends in our two samples, by comparing the distribution of the galaxies residing in voids, sheets, 
filaments and knots \citep[for the exact definition of the environment classification see][]{Eardley2015}, and we do not see a large difference (see Figure \ref{fig:distributions}). It should be noted that the classification of galaxies in \citet{Eardley2015} is only evaluated up to redshift $z = 0.263$, and because of that this test is only indicative. 

\begin{table}
	\caption{Overview of median stellar masses of central galaxies, median redshifts and number of lenses in each selected sample. Stellar masses are taken from version 16 of the stellar mass catalogue, an updated version of the catalogue created by \citet{Taylor2011}.}
	\label{tab:sample_properties}
	\centering
	\begin{tabular}{lccr} 
		\toprule
		Sample & $\log\left(\langle M_{\star} /[M_{\odot} h^{-1}] \rangle \right)$ & $\langle z \rangle $ & Number of lenses\\
		\midrule
		\emph{Full}		& 11.32 & 0.188 & 2061\\
		$\langle R \rangle^{+}$ & 11.33 & 0.186 & 987\\
		$\langle R \rangle^{-}$ & 11.30 & 0.190 & 1074\\
		\bottomrule
	\end{tabular}
\end{table}
 
\subsection{Measurement of the ESD profile}
\label{sec:measurement}

We use imaging data from $180$ deg$^2$ of the Kilo-Degree Survey \citep[KiDS;][]{Kuijken2015, DeJong2015} that overlaps with the GAMA survey \citep{Driver2011}, to obtain shape measurements of the galaxies.
KiDS is a four-band imaging survey conducted with the OmegaCAM CCD mosaic camera 
mounted at the Cassegrain focus of the VLT Survey Telescope (VST); the camera and telescope combination provides us with a fairly uniform point spread function 
across the field-of-view.

From the KiDS data we use the $r$-band based shape measurements of galaxies, with an average seeing of $0.66$ arcsec. The image reduction, photometric redshift calibration and shape measurement 
analysis is described in detail in \citet{Hildebrandt2016}.

We measure galaxy shapes using \emph{lens}fit \citep[][where the method calibration is described]{Miller2013, Conti2016},
which provides measurements of the galaxy ellipticities ($\mathrm{\epsilon_{1}}$, $\mathrm{\epsilon_{2}}$) with respect to an equatorial coordinate system. For each source-lens pair we compute the tangential $\epsilon_{\rm t}$ and cross component $\epsilon_{\times}$ of the source's ellipticity around the position of the lens:
\begin{equation}
\begin{bmatrix} \epsilon_{\rm t} \\
\epsilon_{\times}
\end{bmatrix}= \begin{bmatrix} -\cos(2\phi) & -\sin(2\phi) \\
\phantom{-}\sin(2\phi) & -\cos(2\phi) 
\end{bmatrix} \begin{bmatrix} \epsilon_{1} \\
\epsilon_{2}
\end{bmatrix}
,
\end{equation}
where $\mathrm{\phi}$ is the angle between the $x$-axis and the lens-source separation vector. 

The azimuthal average of the tangential ellipticity of a large number of galaxies in the same area of the sky is an unbiased estimate of the shear. 
On the other hand, the azimuthal average of the cross ellipticity over many sources should average to zero \citep{Schneider2003}. 
Therefore, the cross ellipticity is commonly used as an estimator of possible systematics in the measurements such as non-perfect PSF deconvolution, centroid bias and pixel level detector effects. 
Each lens-source pair is then assigned a weight
\begin{equation}
\label{eq:weights}
\tilde{w}_{\mathrm{ls}}=w_{s} \left(\tilde \Sigma_{\mathrm{cr, ls}}^{-1}\right)^{2} \, ,
\end{equation}
which is the product of the \emph{lens}fit weight $w_{s}$ assigned
to the given source ellipticity and $\tilde\Sigma_{\mathrm{cr, ls}}^{-1}$ -- the effective inverse critical surface mass density, 
which is a geometric term that downweights lens-source pairs that are close in redshift.
We compute the effective inverse critical surface mass density for each lens using the spectroscopic redshift of the lens $z_{\mathrm{l}}$ and the full 
redshift probability distribution of the sources, $n(z_{\mathrm{s}})$, calculated using a direct calibration method presented in \citet{Hildebrandt2016}.
This is different from what was presented in \citet{Viola2015} and used in previous studies on KiDS DR1/2 data, where they used individual $p(z_{\mathrm{s}})$ per source galaxy. 
The effective inverse critical surface density can be written as:
\begin{equation}
\label{eq:crit_effective}
\tilde\Sigma_{\mathrm{cr, ls}}^{-1}=\frac{4\pi G}{c^2} D(z_{\mathrm{l}}) \int_{z_{\mathrm{l}} + \delta_{z}}^{\infty} \frac{D(z_{\mathrm{l}},z_{\mathrm{s}})}{D(z_{\mathrm{s}})}n(z_{\mathrm{s}})\, \mathrm{d}z_{\mathrm{s}} \, ,
\end{equation}
where $\delta_{z}$ is an offset to mitigate the effects of contamination from the group galaxies (see Appendix \ref{ref:systematics}).
We determine the $n(z_{\mathrm{s}})$ for every lens redshift separately, by selecting all galaxies in the spectroscopic sample with a $z_{s}$ larger than $z_{\mathrm{l}} +\delta_{z}$, with $\delta_{z} = 0.2$. The same cut is applied to the photometric redshifts $z_{\mathrm{s}}$ of the sources entering the calculation of the lensing signal.
This condition was not necessary in \citet{Viola2015} as the individual $p(z_{\mathrm{s}})$ accounted for the possible cases when the sources would be in front of the lens. 
Thus, the ESD can be directly computed (using Equation \ref{eq:deltaSigmaTheo}) in bins of projected distance $R$ to the lenses as:
\begin{equation}
\label{eq:ESDmeasured}
\Delta \Sigma (R)=\left[ \frac{\sum_{\mathrm{ls}}\tilde{w}_{\mathrm{ls}}\epsilon_{\mathrm{t, s}}\Sigma_{\mathrm{cr, ls}}^{\prime}}{\sum_{\mathrm{ls}}\tilde{w}_{\mathrm{ls}}} \right] \frac{1}{1+\mu} \, .
\end{equation}
where $\Sigma_{\mathrm{cr, ls}}^{\prime} \equiv 1/ \tilde\Sigma_{\mathrm{cr, ls}}^{-1}$ and the sum is over all source-lens pairs in the distance bin, and
\begin{equation}
\mu = \frac{\sum_{\rm i}w_{\rm i}^{\prime}m_{\rm i}}{\sum_{\rm i}w_{\rm i}^{\prime}} \, ,
\end{equation}
is an average correction to the ESD profile that has to be applied to correct for the multiplicative bias $m$ in the \emph{lens}fit shear estimates. 
The sum goes over thin redshift slices for which $m$ is obtained using the method presented in \citet{Conti2016}, weighted 
by $w^{\prime} = w_{\rm s}D(z_{\rm l},z_{\rm s}) / D(z_{\rm s})$ for a given lens-source sample. The value of $\mu$ is 
around $- 0.014$, independent of the scale at which it is computed. Estimates of $m$ for each 
redshift slice used in the calculation are presented in Figure \ref{fig:m_slices}.

It should be noted that the photometric redshift calibration and shape 
measurement steps differ significantly from the methods used in \citet{Viola2015} and thus we have 
to examine the possible systematic errors and biases. In order to do so, 
we devise a number of tests to see how the data behave in different observational limits, 
and the results are presented in Appendix \ref{ref:systematics}. We test for the presence of additive bias as well as for the presence of cross shear over a wide range of scales. Furthermore, we check how much the GAMA galaxy group members contaminate our source population, and what differences are introduced by the use of a global $n(z_{\rm s})$ instead of individual $p(z_{\rm s})$ per galaxy. We conclude that one should use comoving scales between $70$ kpc/$h$ and $10$ Mpc/$h$ (this range is motivated by the significant contamination by the GAMA group galaxies on the source population on small scales, and non-vanishing cross-term and additive biases present in the lensing signal calculated around random points on large scales), and use between $5$ and $20$ 
radial bins, depending on the choice of error estimation technique and the maximum scale, which is dictated by the number of independent regions one can use to estimate the bootstrap errors and the number of independent entries in the resulting covariance matrix (see further motivation in Section \ref{sec:covariance}). Here, we use 8 radial bins between $70$ kpc/$h$ and $10$ Mpc/$h$. For the sources we adopt the redshift range $[0.1, 0.9]$, motivated by \citet{Hildebrandt2016}.

\subsection{Covariance matrix estimation}
\label{sec:covariance}

\begin{figure}
	\centering
	\includegraphics[width=0.9\columnwidth]{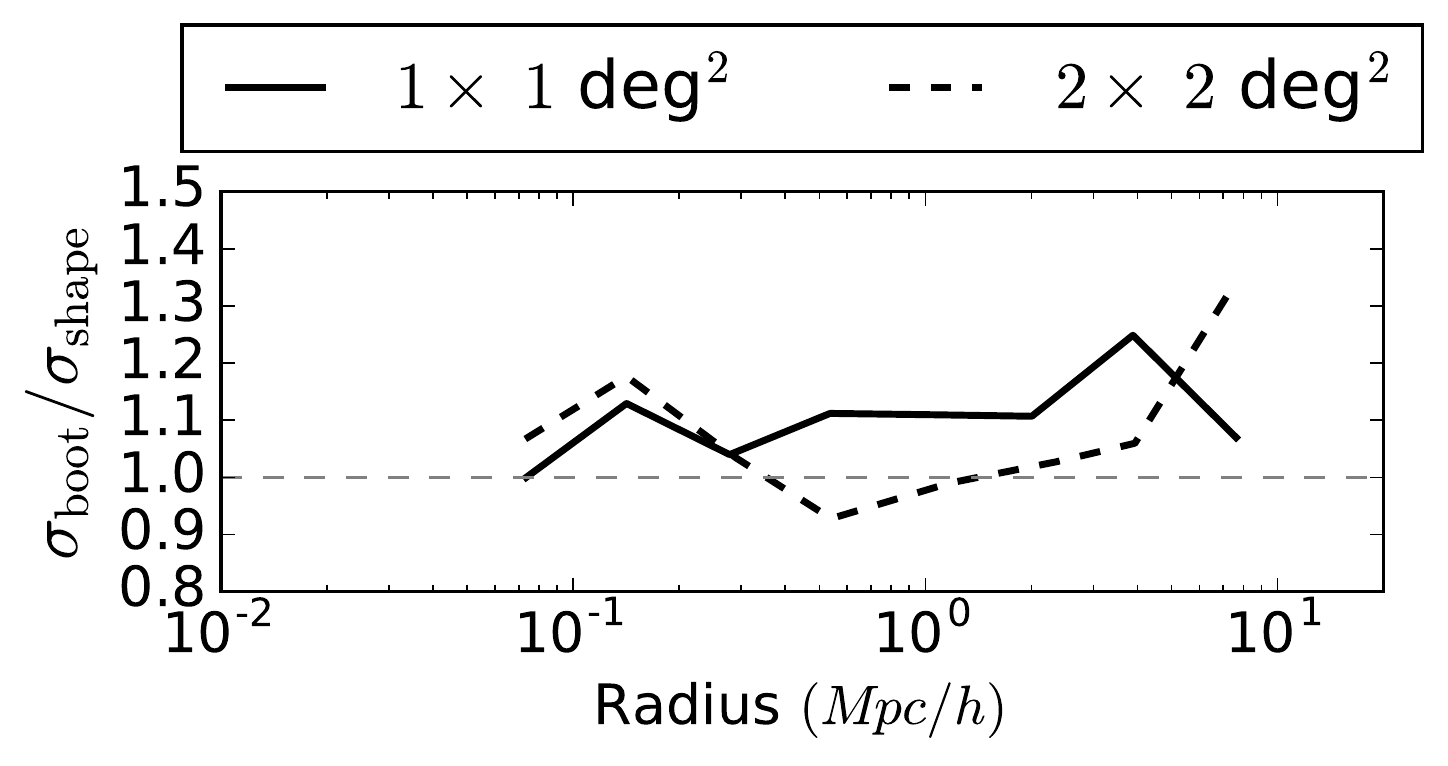}
 	\caption{Ratios of the errors obtained using a bootstrap method and the 
	errors obtained from the analytical covariance. Rations for $1$ deg$^{2}$ KiDS tiles and $4$ deg$^{2}$ patches are shown in solid and dashed black lines. The errors are taken as the square root of the diagonal of the 
	respective covariance matrices.}
	\label{fig:covariance_comparison}
\end{figure}

Statistical error estimates on the lensing signal are obtained in two ways. First we follow the prescription 
used in \citet{Viola2015} which was shown to be valid in \citet{Sifon2015}, \citet{VanUitert2016} and \citet{Brouwer2016}, where we 
calculate the analytical covariance matrix from the contribution of each source in radial bins. This prescription accounts for shape noise of source galaxies and includes information about the survey geometry (including the masking of the lens and source galaxies). However, this method does not account for sample variance, but \citet{Viola2015} showed that this prescription works sufficiently well up to $2$ Mpc/$h$. As we calculate 
the lensing signal up to $10$ Mpc/$h$, we use the bootstrap method, as the analytical 
covariance tends to underestimate the errors on scales greater than $2$ Mpc/$h$ (see Figure \ref{fig:covariance_comparison}, where we compare the different methods for estimating the errors). We first test the bootstrap method by bootstrapping the lensing signal measured around lenses in 
each of the $1$ deg$^{2}$ KiDS tiles. We randomly select $180$ of these tiles with replacement and stack the signals. We repeat this procedure $10^{5}$ times. The covariance matrix is well constrained by the $180$ KiDS tiles used in this analysis, as the number of independent entries in the covariance matrix is equal to $36$.

\begin{figure}
	\centering
	\includegraphics[width=\columnwidth]{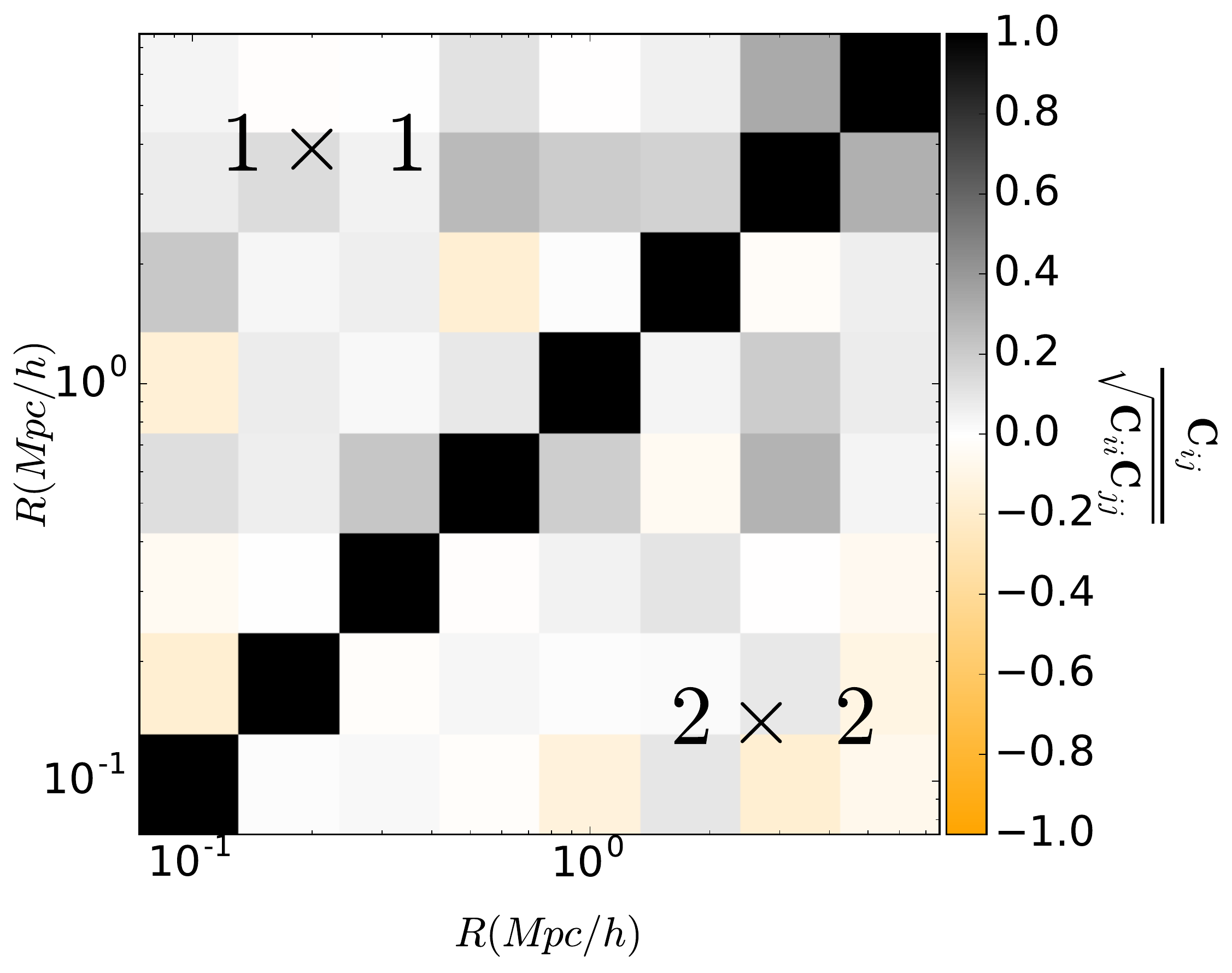}
 	\caption{The ESD correlation matrix between different radial bins estimated using a bootstrap technique. Bootstrap covariance accounts both for shape noise and cosmic variance. In the upper triangle we show the correlation matrix when using $1$ deg$^{2}$ tiles, and in the lower triangle the correlation matrix when using $4$ deg$^{2}$ patches (as indicated).}
	\label{fig:covariance_matrix}
\end{figure}

As the physical size of the tile is comparable to the maximum separations we are considering (one degree at the median redshift of our sample corresponds to \mbox{\textasciitilde{} $8$ Mpc/$h$}), there is a concern that the KiDS tiles might not well describe the errors on scales larger than $2$ Mpc/$h$, because the tiles are not truly independent from each other. In fact, the sources in neighbouring tiles do contribute to the lensing signal of a group in a certain tile and the tiles are thus not independent on scales above $8$ Mpc/$h$. We thus repeat the above exercise and calculate the bootstrapped covariance matrix using $4$ deg$^{2}$ KiDS patches (by combining $4$ adjacent KiDS tiles), which leaves us with $45$ independent bootstrap regions (which is still enough to constrain the $36$ independent entries in our covariance matrix). The square root of diagonal elements compared to the result of the analytical covariance can be seen in Figure \ref{fig:covariance_comparison} and the full bootstrap correlation matrix in Figure \ref{fig:covariance_matrix}. For a shape noise dominated measurement one would expect that all three methods yield the same results on scales smaller than $2$ Mpc/$h$. While this holds for all methods on small scales, it certainly does not hold at scales larger than $2$ Mpc/$h$ for the analytical and bootstrap covariances, when taking only $1$ deg$^{2}$ tiles. The main issue here is that one lacks large enough independent regions to properly sample the error distribution on large scales, and thus the resulting errors are highly biased. Taking all this considerations into account, we decide to use the bootstrapping over $4$ deg$^{2}$ patches as our preferred method of estimating the errors of our lensing measurements.

Due to noise, the inverse covariance matrix calculated from the covariance matrix, $\mathbfss{C}^{-1}_{*}$, is not an unbiased estimate of the true inverse covariance matrix 
$\mathbfss{C}^{-1}$ \citep{Hartlap2007}. In order to derive an unbiased estimate of the inverse covariance we need to apply a correction so that 
$\mathbfss{C}^{-1} = \alpha\, \mathbfss{C}^{-1}_{*}$. In the case of Gaussian errors and statistically independent data vectors, this correction factor is:
\begin{equation}
\alpha = \frac{n-p-2}{n-1}\, ,
\end{equation}
where $n$ is the total number of independent bootstrap patches, i.e. $45$ in our case, and $p$ is the number of data points we use, i.e. in our case $8$. \citet{Hartlap2007} also show that for $p/n \lesssim 0.8$ (in our case we have $p/n = 0.18$) this correction produces an unbiased estimate of the inverse covariance matrix $\mathbfss{C}^{-1}$ and we use this correction in our analysis.

When fitting the halo model to the data, we use the inverse covariance matrix from the bootstrap using $4$ deg$^{2}$ patches. One could use more sophisticated methods to precisely estimate the errors on very large scales. For instance, the analytical covariance method from \citet{Hildebrandt2016} can be adapted for galaxy-galaxy lensing or using galaxy-galaxy lensing specific mock catalogues to estimate the covariance matrix. Future studies using the KiDS data, expanding the analysis over greater separations or simply having more data points should employ methods like that one, but for the purposes 
of this study, the covariance matrix presented here is sufficient.

\section{Halo model}
\label{sec:halomodel_intro}

A successful analytic framework to describe the clustering of dark matter and its evolution in the Universe  is 
the halo model \citep{Seljak2000, Peacock2000, Cooray2002, Bosch2012, Mead2015}. The halo model provides 
an ideal framework to describe the statistical weak lensing signal around a selection of galaxies. One of the assumptions of the halo model is that halo bias is only a function of halo mass, an assumption we want to test in this work. The halo model is built upon the statistical description 
of the properties of dark matter haloes (namely the average density profile, large scale bias and abundance) as well as on the 
statistical description of the galaxies residing in them. 

The mass of a dark matter halo in the halo model framework is defined as:
\begin{equation}
\label{eq:over_dens_mass}
M = \frac{4\pi}{3}r^{3}_{\Delta}\Delta\overline{\rho}_{\text{m}} \,,
\end{equation}
enclosed by the radius $r_{\Delta}$ within which the 
mean density of the halo is $\Delta$ times $\overline{\rho}_{\text{m}}$. Throughout the paper we use 
$\overline{\rho}_{\text{m}}$ as the mean comoving matter density of the Universe 
\mbox{($\overline{\rho}_{\text{m}} = \Omega_{\text{m}, 0} \, \rho_{\text{crit}}$}, 
where \mbox{$\rho_{\text{crit}} = 3H^{2}_{0}/8\pi G$} and $\Delta = 200$). We assume that the density profile of dark matter haloes follows an NFW profile \citep{Navarro1997}.

\subsection{Model specifics}
\label{sec:halomodel}

The ESD profile as defined in Equation \ref{eq:ESD}, which is related to 
the galaxy-matter cross-correlation function $\xi_{\text{g,m}}(r, z)$, can be obtained by Fourier transforming the galaxy-matter 
power spectrum $P_{\text{g,m}}(k, z)$:
\begin{equation}
\label{eq:lens_fourier}
\xi_{\text{g,m}}(r, z) = \frac{1}{2\pi^{2}} \int^{\infty}_{0} P_{\text{g,m}}(k, z) \frac{\sin kr}{kr} k^2\, \mathrm{d}k \,,
\end{equation}
where $k$ is the wavenumber and the subscripts m and g stand for \emph{matter} and \emph{galaxy}. Equation \ref{eq:lens_fourier} 
can be expressed as a sum of a term that describes the small scales (one-halo, 1h), and one describing 
the large scales (two-halo, 2h) (see Equation \ref{eq:power1}). 

As we calculate the stacked ESD profile around the central galaxies of the GAMA groups, the only contribution to 
the one-halo term arises from central galaxies. The contribution of satellite galaxies is not modelled as it does not induce coherent distortions in our stacked measurements. As galaxies are not isolated at large scales, the signal there is dominated by the clustering of dark matter halos. This so-called two-halo term 
will play an important role in characterising halo assembly bias. Thus, we write the power spectrum
as:
\begin{equation}
\label{eq:power1}
P_{\text{g,m}}(k, z) = P_{\text{g,m}}^{1\text{h},c}(k, z) + P_{\text{g,m}}^{2\text{h},c}(k, z) \,,
\end{equation}
where:
\begin{equation}
\label{eq:1halo1}
P_{\text{g,m}}^{1\text{h},c}(k, z) =  \frac{1}{\overline{\rho}_{\text{m}} \overline{n}_{\text{g}}} \int \mathrm{d}M\frac{\mathrm{d}n(M, z)}{\mathrm{d}\ln M} u_{\text{g}}(k\vert M)\langle N_{\text{g}}^{c}\vert M \rangle \,,
\end{equation}
and $\frac{\mathrm{d}n(M, z)}{\mathrm{d}\ln M}$ is the halo mass function (number density of 
haloes as a function of their mass), $\langle N_{\text{g}}^{c}\vert M \rangle$ is an average number 
of central galaxies residing in a halo with given mass $M$ and the $u_{\text{g}}(k\vert M)$ is the normalised Fourier transform of the group density profile. For the halo mass function we use the analytical 
function presented in \citet{Tinker2010}. Furthermore we define the comoving number density of 
groups $\overline{n}_{\text{g}}$ as:
\begin{equation}
\label{eq:HOD3}
\overline{n}_{\text{g}} = \int \langle N_{\text{g}}^{c}\vert M \rangle \frac{\mathrm{d}n(M, z)}{\mathrm{d}\ln M} \frac{\mathrm{d}M}{M} \,.
\end{equation}

We require that the halo mass function obeys the following normalization relation:
\begin{equation}\label{eq:dens1}
\int^{\infty}_{0}  \mathrm{d}M \frac{\mathrm{d}n(M, z)}{\mathrm{d}\ln M} = \overline{\rho}_{\text{m}} \,,
\end{equation}
which is satisfied in the case of using the halo mass function from \citet{Tinker2010}. The two-halo term can be written as:
\begin{equation}
\label{eq:2halo1}\
P_{\text{g,m}}^{2\text{h},c}(k, z) = b \, P_{\text{m}}(k, z) \,,
\end{equation}
where $b = A_{\text{b}} \, b_{\text{g}}$ and $b_{\text{g}}$ is given by:
\begin{equation}
\label{eq:2halo2}
b_{\text{g}} = \frac{1}{\overline{n}_{\text{g}}} \int \langle N_{\text{g}}^{c}\vert M \rangle b_{\text{h}}(M, z) \frac{\mathrm{d}n(M, z)}{\mathrm{d}\ln M} \frac{\mathrm{d}M}{M} \,,
\end{equation}
where $A_{\text{b}}$ is a free parameter that we fit for, $b_{\text{h}}(M,z)$ is the halo bias function and 
$P_{\text{m}}(k, z)$ is the linear matter-matter power spectrum. For the halo bias function 
we use the fitting function from \citet{Tinker2010}, as it was obtained 
using the same numerical simulation from which the halo mass function was calibrated. This form of the two-halo term is motivated by the fact that the halo density contrast and matter density contrast can be related with a halo bias function that can be linearised \citep{Bosch2012}. The extra free parameter $A_{\text{b}}$ is introduced, because any signature of halo assembly bias will break the mass-only Ansatz of the halo model precisely at this point.

We have adopted the parametrization of the concentration-mass relation, given by \citet{Duffy2011}:
\begin{equation}
\label{eq:con_duffy}
c(M, z) = f_{c} \times 10.14\ \left[\frac{M}{(2\times 10^{12} M_{\odot}/h)}\right]^{- 0.081}\ (1+z)^{-1.01} \,,
\end{equation}
with a free normalisation $f_{c}$.

The halo occupation statistics of central galaxies are defined via the function $\langle N_{\text{g}}^{c}\vert M \rangle$, 
the average number of galaxies as a function of halo mass $M$. 
We model $\langle N_{\text{g}}^{c}\vert M \rangle$ as a error function characterised by a minimum mass, ${{\rm log}[{M_{1}}/(h^{-1} M_{\odot})]}$, 
and a scatter $\sigma_{c}$:
\begin{equation}
\label{eq:phi_c}
\langle N_{\text{g}}^{c}\vert M \rangle =
\frac{1}{2} \left[1 + {\rm erf} \left(\frac{\log{M}-\log{M_{1}}}{\sigma_{c}} \right)
\right] \,.
\end{equation} 
We caution the reader against over-interpreting the physical meaning of this parametrisation. This functional form mainly serves the purpose of assigning a distribution of halo masses around a mean halo mass value.

As in \citet{Viola2015} we assume that the degree of miscentering of the groups in three dimensions is proportional to the halo scale radius $r_s$, a function of halo mass and redshift, 
and we parametrise the probability that a central galaxy is
miscentered as $p_{\rm off}$. This gives
\begin{equation}
u_{\text{g}}(k\vert M) = 
u_{\text{m}}(k\vert M)  \, 
\left(1-p_{\rm off} + p_{\rm off} \,
{\mathrm e}^{\left[-0.5 k^2 (r_s {\cal R}_{\rm off})^2\right]}
\right)
\, ,
\end{equation}
where $u_{\text{m}}(k\vert M)$ is the Fourier transform of the normalised dark matter density profile, which is assumed to follow an NFW profile \citep{Navarro1997}, and ${\cal R}_{\rm off}$ the typical miscentering distance.

We include the contribution of the stellar mass of the BCGs to the lensing signal as a point mass approximation, which we can write as:
\begin{equation}
\label{eq:point_mass}
\Delta \Sigma_{\text{pm}} = \frac{\langle M_{\star} \rangle}{\pi R^{2}} \,,
\end{equation}
where $\langle M_{\star} \rangle$ is the average stellar mass of the selected galaxies obtained directly from the GAMA catalogue. Stellar masses are taken from version 16 of the stellar mass catalogue, an updated version of the catalogue created by \citet{Taylor2011}, who fitted \citet{Bruzual2003} synthetic stellar spectra to the broadband SDSS photometry assuming a \citet{Chabrier2003} IMF and a \citet{Calzetti2000} dust law. This stellar mass contribution is kept fixed for all of our samples.

The free model parameters for each sample are ${\lambda = [f_{c}\,, p_{\text{off}}\,, {\cal R}_{\rm off}\,, \log(M_{1})\,, \sigma_{c}\,, b]}$, and when 
fitting we also store the derived parameter $\log(M_{h})$ -- an effective mean halo mass:
\begin{equation}
\label{eq:mass_halo}
M_{\text{h}} =  \frac{1}{\overline{n}_{\text{g}}} \int \langle N_{\text{g}}^{c}\vert M \rangle \frac{\mathrm{d}n(M, z)}{\mathrm{d}\ln M} \mathrm{d}M \,,
\end{equation}
which accounts for weighting of the given fitted masses by the halo mass function. We use this mean halo mass when reporting our results.

\subsection{Fitting procedure}
\label{sec:sampler}

We fit this model to each of our two samples ($\langle R \rangle^{+}$ and $\langle R \rangle^{-}$) 
with independent parameters and covariance matrices. This gives us a total of 12 free parameters. 
We use a Bayesian inference method in order to obtain full posterior 
probabilities using a Monte Carlo Markov Chain (MCMC) technique; more specifically we use the \texttt{emcee} 
Python package \citep{Foreman-Mackey2012}. The likelihood $\mathcal{L}$ is given by
\begin{equation}\label{eq:likelihood}
\mathcal{L} \propto \exp\left[- \frac{1}{2}(\boldsymbol{O}_{i}-\boldsymbol{M}_{i})^{T}\mathbfss{C}^{-1}_{ij}(\boldsymbol{O}_{j}-\boldsymbol{M}_{j})\right] \,,
\end{equation}
where $\boldsymbol{O}_{i}$ and $\boldsymbol{M}_{i}$ are the measurements and model predictions in 
radial bin $i$, $\mathbfss{C}^{-1}_{ij}$ is the element of the inverse 
covariance matrix that accounts for the correlation between radial bins $i$ and $j$. In the fitting procedure 
we use the inverse covariance matrix as described in Section \ref{sec:covariance}. 
We use wide flat priors for all the parameters, and the ranges can be seen in Table \ref{tab:results}. The halo model (halo mass function and the power spectrum) is evaluated at the median redshift for each sample. 
We run the sampler using $120$ walkers, each with $2000$ steps (for the combined number of $240\,000$ samples), 
out of which we discard the first $600$ burn-in steps ($72\,000$ samples). The resulting MCMC chains are well converged according to the integrated autocorrelation time test. 

\begin{figure*}
	\includegraphics[width=0.75\textwidth]{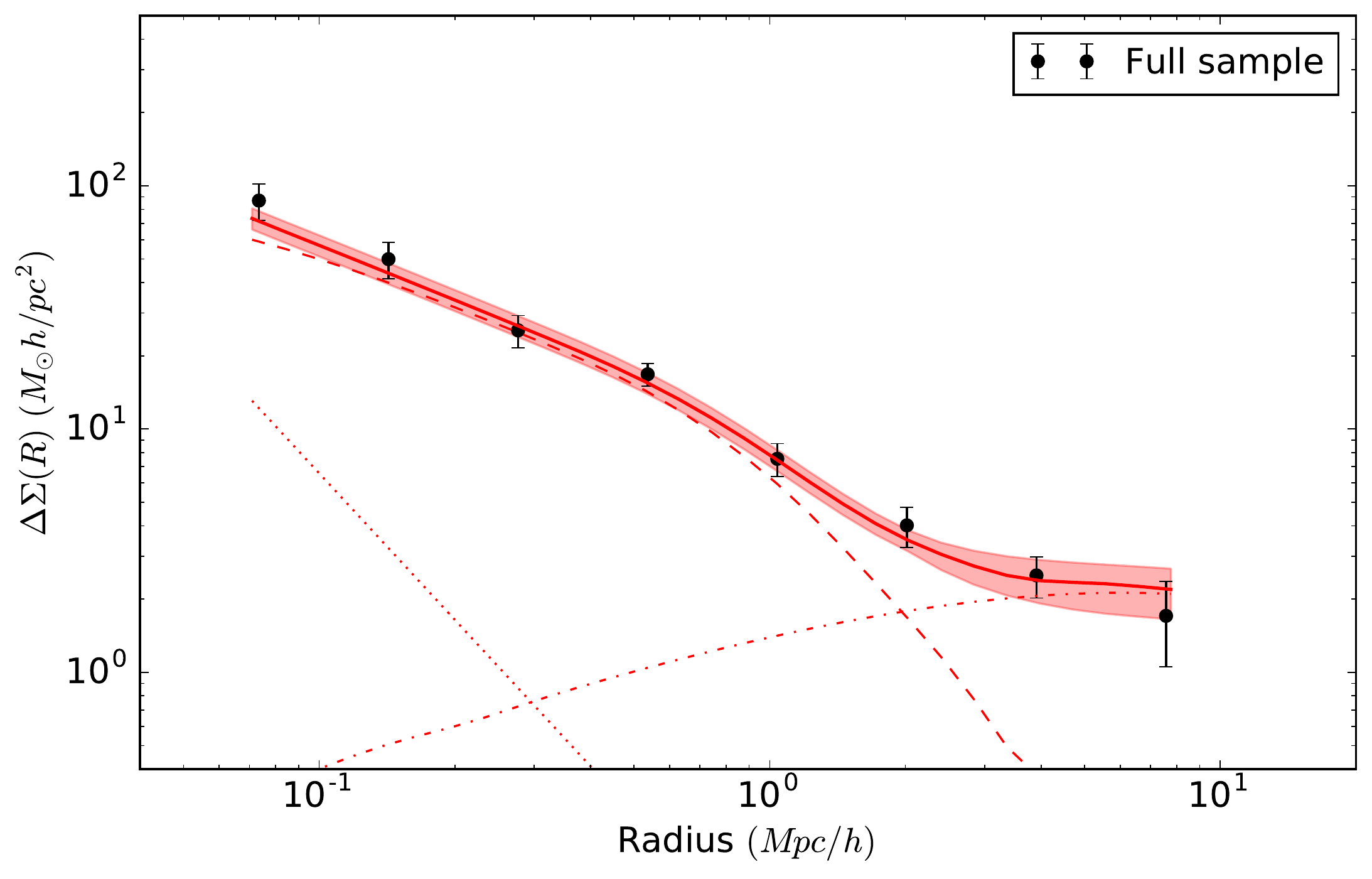}
 	\caption{Stacked ESD profiles measured around the central galaxies of GAMA groups from the full sample of galaxies used in this study. 
	The solid red lines represent
	the best-fitting halo model as obtained using a MCMC fit, with the $68 \%$ confidence interval indicated with a shaded region. 
	Dashed, dash-dotted and dotted lines represent the one-halo term, 
	two-halo term and stellar contribution, respectively (see Section \ref{sec:halomodel}).}
	\label{fig:ESD_full}
\end{figure*}

Figure \ref{fig:ESD_full} shows the stacked ESD profile for all $2061$ galaxy groups (full sample). In comparison to \citet{Viola2015}, this sample has around \mbox{\textasciitilde{} $40 \%$} more galaxy groups, given by the fact we are using the full equatorial KiDS and GAMA overlap. We calculate the lensing signal for all our samples according to the procedure described in Section \ref{sec:measurement}. In the same figure, we also show the halo model fit to the data, as described in this section.

\section{Results}
\label{sec:results}

\label{sec:res_lensing}

\begin{figure*}
	\centering
	\includegraphics[width=0.75\textwidth]{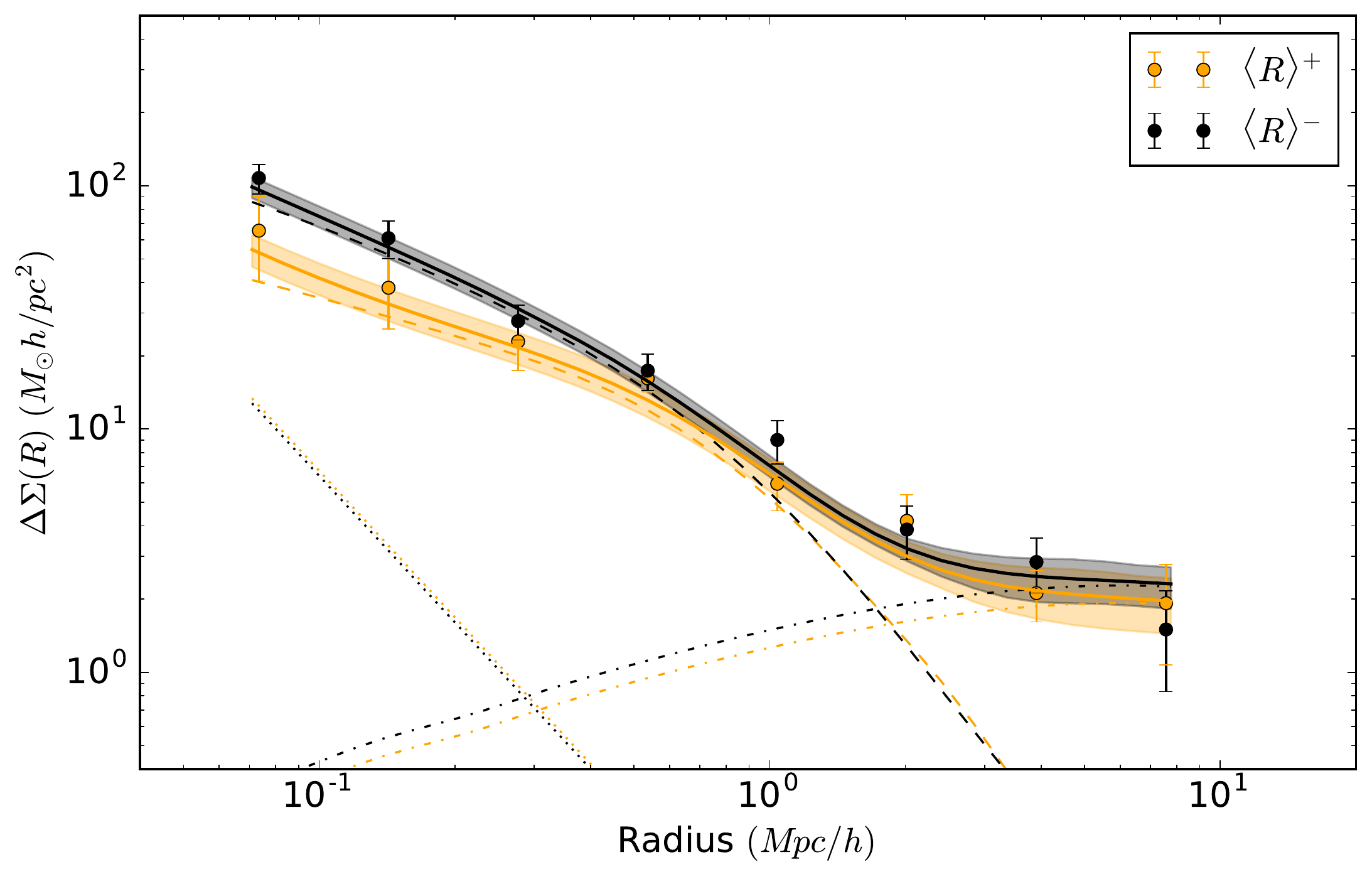}
 	\caption{Stacked ESD profiles measured around the central galaxies of GAMA groups, selected according to the 
	average separation of satellite galaxies (see Section \ref{sec:lenses}). The solid orange and black lines represent 
	the best-fitting halo model as obtained using a MCMC fit, with the $68 \%$ confidence interval indicated with a shaded region. 
	Dashed, dash-dotted and dotted lines represent the one-halo term, 
	two-halo term and stellar contribution, respectively.}
	\label{fig:ESD}
\end{figure*}

We fit the halo model as presented in Section \ref{sec:halomodel} to the two subsamples ($\langle R \rangle^{+}$ -- sample with more dispersed satellite galaxies and $\langle R \rangle^{-}$ -- sample with more concentrated satellite galaxies). 
The fits have a reduced $\chi^{2}_{\text{red}}$ ($=\chi^{2}/\text{d.o.f}$) equal to 1.31 and 1.41 for the $\langle R \rangle^{+}$ and $\langle R \rangle^{-}$ sample, respectively, and the best fit models are presented in Figure \ref{fig:ESD}, plotted with the $16$ and $84$ percentile 
confidence intervals. We also plot the stacked ESD profiles for both samples of galaxies, with $1\sigma$ error bars, 
which are obtained by taking the square root of the diagonal elements of the bootstrap covariance matrix. 

\begin{figure}
	\centering
	\includegraphics[width=\columnwidth]{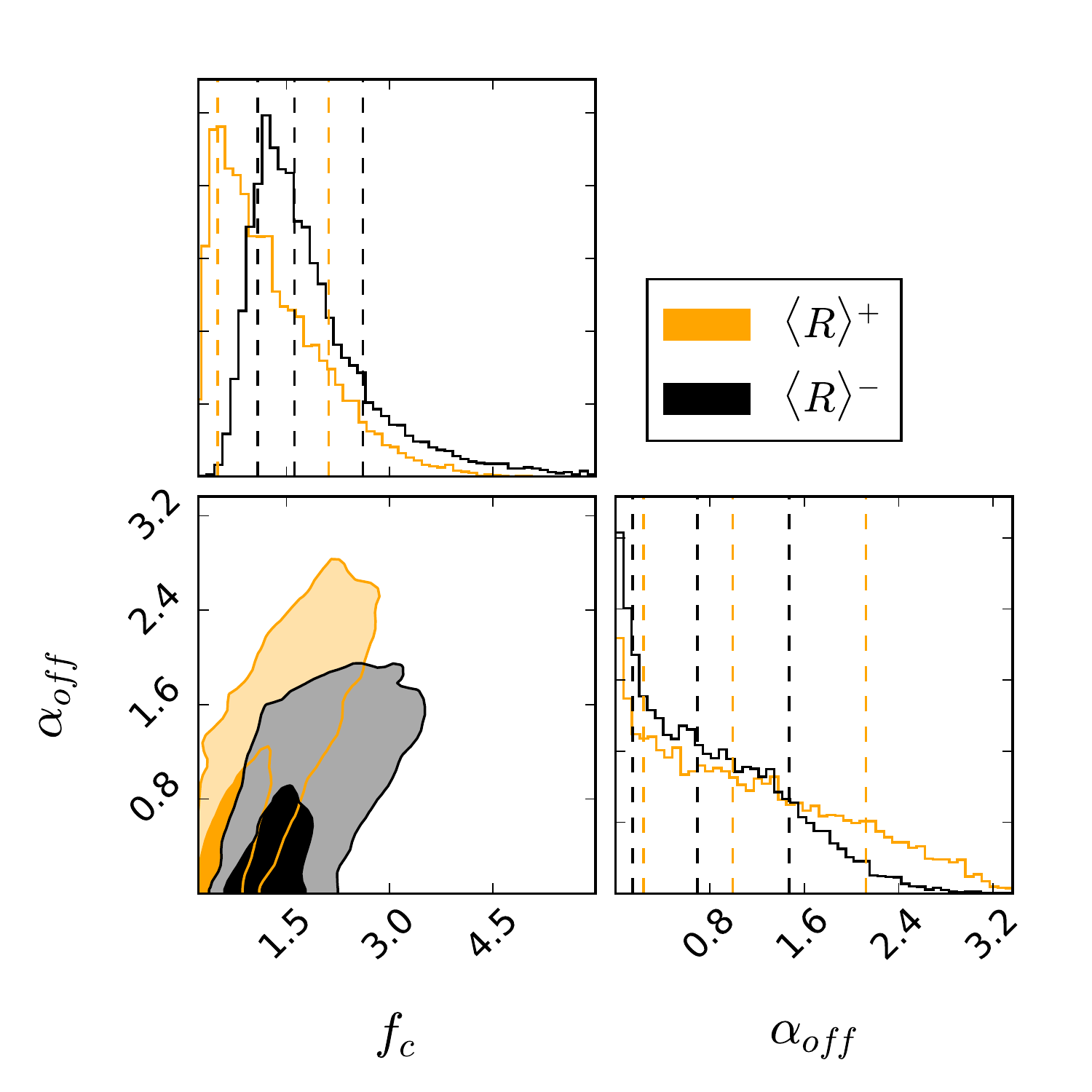}
 	\caption{The posterior distributions of the average projected offset $\alpha_{\text{off}}$ 
	and the normalisation of the concentration-halo mass relation $f_{c}$. The contours indicate $1\sigma$ and 
	$2\sigma$ confidence regions.}
	\label{fig:corner_miscenter}
\end{figure}

\begin{figure}
	\centering
	\includegraphics[width=\columnwidth]{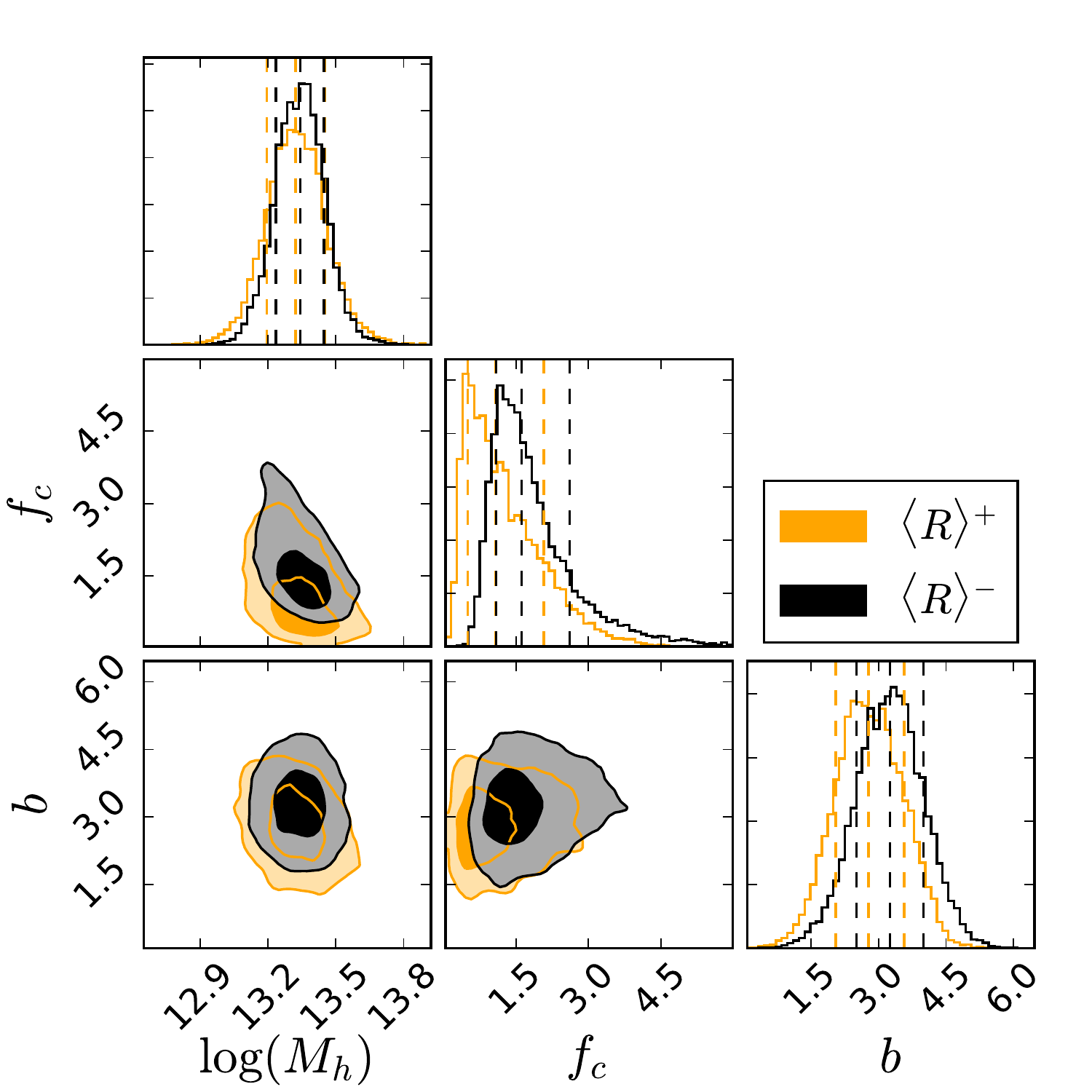}
 	\caption{The posterior distributions of the halo model parameters $M_{h}$, $f_{c}$ and $b$. 
	The posterior distributions clearly show a slight difference in the obtained halo masses as well as no 
	difference in the obtained halo biases. The contours indicate $1\sigma$ and $2\sigma$ confidence regions.}
	\label{fig:corner_bias}
\end{figure}

\begin{figure}
	\centering
	\includegraphics[width=\columnwidth]{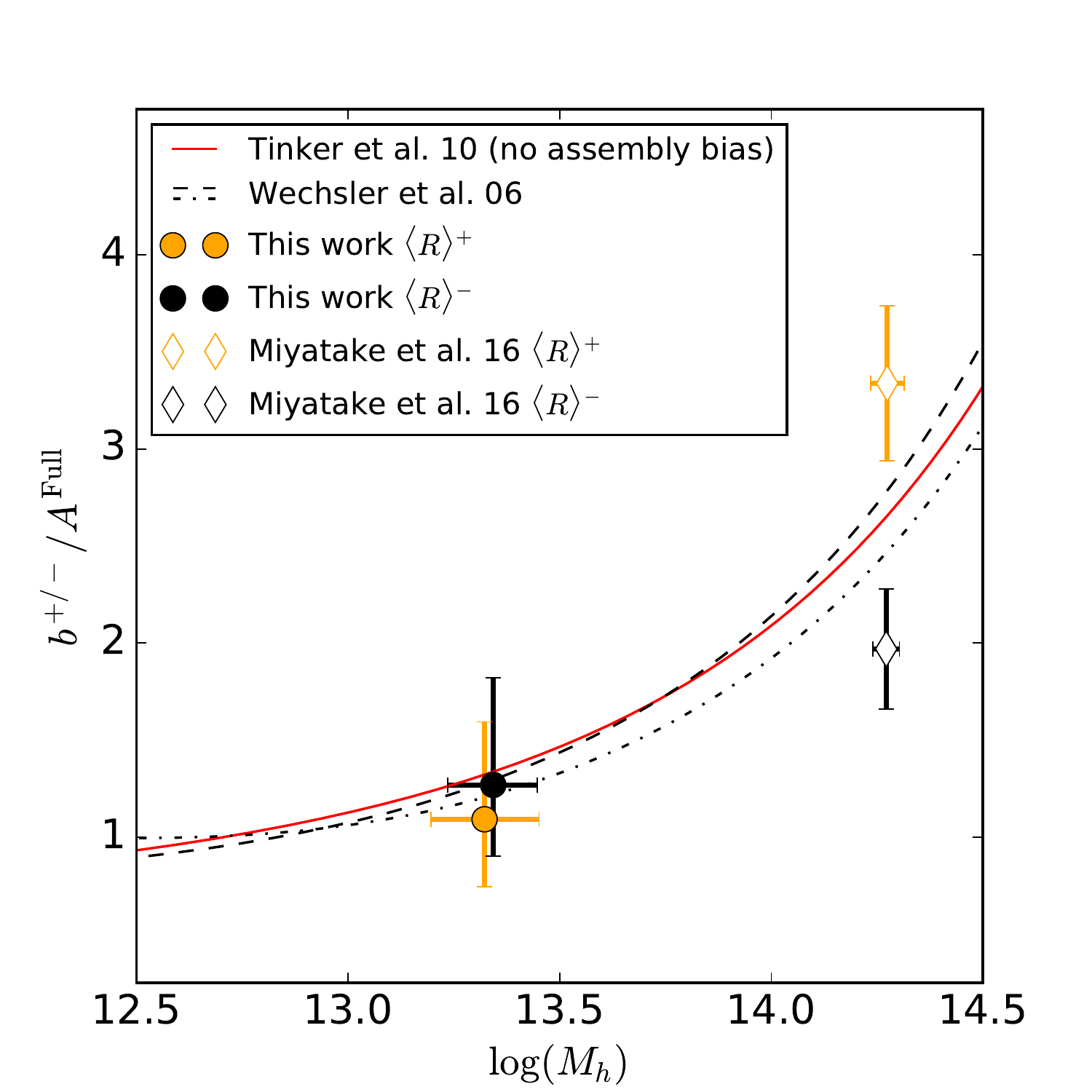}
 	\caption{Comparison between the halo bias $b$ and the predictions from the halo bias function from \citet{Tinker2010} and the concentration dependent halo bias from \citet{Wechsler2006}, as a function of halo mass $M_{h}$. 
	Here circles with error bars show the best fit value for $b$ for each sample and diamonds show the results from \citet{Miyatake2015}. 
	The halo bias function from \citet{Tinker2010} is shown with a red line and the predictions from \citet{Wechsler2006} for different values of $c'$ and a halo collapse mass $M_{c} = 2.1 \times 10^{12} M_{\odot} /h$ (as defined therein). The dashed and dash-dotted lines are predictions for $c'$ derived for our two samples, $\langle R \rangle^{+}$ and 
$\langle R \rangle^{-}$, respectively. Note that the biases are normalised by the $A^{\text{full}}$.} 
	\label{fig:corner_intrinsic}
\end{figure}

\begin{table*}
	\caption{Summary of the lensing results obtained using MCMC halo model fit to the data. All the parameters are defined in Section \ref{sec:halomodel}. $f_c$ is the normalisation of the concentration-halo mass relation, $p_{\text{off}}$ the miscentering probability, ${\cal R}_{\rm off}$ the miscentering distance, $M_{1}$ central mass used to parametrise the HOD, $\sigma_{c}$ scatter in HOD distribution and $b$ bias.}
	\begin{threeparttable}
	\centering
	\label{tab:results}
	\begin{tabular}{lccccccc} 
		\toprule
		Sample & $\log (M_{h}[M_{\odot} /h])$ & $f_{c}$ & $p_{\text{off}}$ & ${\cal R}_{\rm off}$ & $\log(M_{1}[M_{\odot} /h])$ & $\sigma_{c}$ & $b$\\
		\midrule
		\addlinespace
		Priors & -- & $[0.0, 6.0]$\ & $[0.0, 1.0]$ & $[0.0, 3.5]$ & $[11.0, 17.0]$ & $[0.05, 1.5]$ & $[0.0, 10.0]$\\
		\midrule
		\addlinespace
		$\langle R \rangle^{+}$ & $13.32^{+0.13} _{-0.13}$ & $1.08^{+0.99} _{-0.58}$ & $0.58^{+0.27}_{-0.36}$ & $2.10^{+0.99} _{-1.23}$ & $13.07^{+0.19} _{-0.18}$ & $0.60^{+0.05} _{-0.05}$ & $2.77^{+0.78} _{-0.73}$\\
		\addlinespace
		
		$\langle R \rangle^{-}$ & $13.34^{+0.10} _{-0.11}$ & $1.61^{+0.99} _{-0.53}$ & $0.37^{+0.24}_{-0.23}$ & $2.40^{+0.81} _{-1.50}$ & $13.10^{+0.17} _{-0.16}$ & $0.61^{+0.05} _{-0.05}$ & $3.25^{+0.74} _{-0.74}$ \\ 
		\addlinespace
		\midrule
		\addlinespace
		\emph{Full} & $13.42^{+0.09} _{-0.08}$ & $1.03^{+0.63} _{-0.35}$ & $0.42^{+0.21}_{-0.24}$ & $2.46^{+0.73} _{-1.24}$ & $13.22^{+0.14} _{-0.13}$ & $0.60^{+0.05} _{-0.05}$ & $3.05^{+0.72} _{-0.75}$ \\
		\addlinespace
		\bottomrule
	\end{tabular}
	\end{threeparttable}
\end{table*}

The measured parameters are summarised in Table \ref{tab:results}, and their full posterior distributions are shown in 
Figure \ref{fig:triangle_full}. The various parameters show similar results between the $\langle R \rangle^{+}$ and 
$\langle R \rangle^{-}$ subsamples. The normalisations of the concentration-halo mass relations $f_{c}$ are $f_{c}^{+}= 1.08^{+0.99} _{-0.58}$ and $f_{c}^{-} = 1.61^{+0.99} _{-0.53}$ for $\langle R \rangle^{+}$ and 
$\langle R \rangle^{-}$ respectively, in accordance with the results for the full sample (see Table \ref{tab:results}). 
Furthermore the scatter in halo masses, $\sigma_{c}$ is constrained to \mbox{\textasciitilde{} $0.6$} for both samples and it is also 
consistent with the results for the full sample (see Table \ref{tab:results}). 
We observe lower probabilities for miscentering of the central galaxy than reported in \citet{Viola2015}, but with a larger miscentering distance. 
It should be noted, that the average projected offset $\alpha_{\text{off}}$ ($\alpha_{\text{off}} = p_{\text{off}} \times {\cal R}_{\rm off}$) is highly 
degenerate with the concentration normalisation $f_{c}$ and the posterior probability distribution is shown in Figure \ref{fig:corner_miscenter}. 
The resulting degeneracy is similar to the one presented in \citet{Viola2015}.

Since we consider ESD profiles out to $10$ Mpc/$h$, the halo masses are well 
constrained by the inner-most part of the same ESD profile ($r_{200}$ associated with the this mass scale is significantly smaller than $10$ Mpc/$h$). The contribution 
to the ESD profile beyond $2$ Mpc/$h$ can be associated purely with the two-halo term (see Figure \ref{fig:ESD}). 
The ratio of the obtained halo biases is $b^{+}/b^{-} = 0.85^{+0.37}_{-0.25}$. The posterior probability distributions of the obtained halo masses and biases can be seen in Figures \ref{fig:corner_bias} and \ref{fig:triangle_full}.

With the lensing measurements providing us the same halo masses for the two samples (within the errors), we report a null detection of halo assembly bias on galaxy groups scales.
Our result is in accordance with what one would expect if halo bias is only a function of mass (see Figure \ref{fig:corner_intrinsic}). In Figure \ref{fig:corner_intrinsic}, we also compare our results with the biases obtained by \citet{Miyatake2015} and to the predictions for a concentration dependent halo bias from \citet{Wechsler2006}. To account for the slightly different masses of our two samples one can also compare the difference arising purely from 
the normalisation of the bias $A_{\text{b}}$ (as defined in Equation \ref{eq:2halo1}). The ratio of obtained normalisations is still compatible with a null detection; $A_{\text{b}}^{+} / A_{\text{b}}^{-} = 0.86^{+0.43}_{-0.28}$ \,($0.4\sigma$).

If the halo assembly bias due to different spatial distributions of satellite galaxies traces the halo bias due to different halo concentrations, then one would expect that the halo assembly bias would follow the predictions presented in \citet{Wechsler2006}, and would also not be significant near the halo collapse mass $M_c$. The halo collapse masses for our two samples are $M_c = 2.12 \times 10^{12} M_{\odot} /h$ and $M_c = 2.02 \times 10^{12} M_{\odot} /h$ for the $\langle R \rangle^{+}$ and 
$\langle R \rangle^{-}$ subsamples, which are \mbox{\textasciitilde{}  $8\sigma$} below the obtained halo masses. The cancelation effect of the halo assembly bias due to the predicted sign change (clearly seen in Figure \ref{fig:corner_intrinsic}) of the concentration dependent halo bias near the $M_c$ cannot be the cause of the null detection of halo assembly bias, as none of our lenses have halo masses that are below the $M_c$. We however acknowledge that the differences in predicted halo bias following \citet{Wechsler2006} for $c'$ (as defined therein) of our two samples at the obtained halo masses are rather small (halo bias ratio of $1.06$) and challenging to observe in the first place.

As the results can potentially depend on the choice of the concentration-mass relation, and to see if the choice of our fiducial \citet{Duffy2011} concentration-mass relation does not significantly influence our results, we perform a test where we change the fiducial concentration-mass relation to a parameter that is constant with mass and free to fit. The obtained concentrations for the $\langle R \rangle^{+}$ and 
$\langle R \rangle^{-}$ subsamples are $c^{+}= 5.64^{+3.64}_{-2.57}$ and $c^{-} = 8.36^{+2.38}_{-2.14}$ -- again highly degenerate with the average projected offset $\alpha_{\text{off}}$. The ratio of obtained halo biases in this case is $b^{+}/b^{-} = 0.86^{+0.41}_{-0.28}$ and the ratio of obtained normalisations is  $A_{\text{b}}^{+} / A_{\text{b}}^{-} = 0.89^{+0.45}_{-0.31}$. We further check if the method presented can detect a bias ratio different than unity using a sample which is known to have one. For this we split our full sample into two samples with different apparent richnesses by making a cut at $N_{\text{FoF}} = 10$ (in order to have two samples with comparable $S/N$). We fit the halo model as presented in Section 4.1 to obtain the posterior distributions of the halo biases. As expected, the two samples have significantly different halo masses with the high richness sample having a halo mass of $\log (M_{h}[M_{\odot} /h]) = 13.72^{+0.13}_{-0.11}$ and the low richness sample having a halo mass of $\log (M_{h}[M_{\odot} /h]) = 13.24^{+0.09}_{-0.09}$. The obtained halo bias ratio is, as expected, different than unity $b^{\text{high}}/b^{\text{low}} = 2.84^{+1.75}_{-1.01}$, which is also true when one accounts for the fact that the samples have different halo masses. In this case, the ratio of obtained normalisations is $A_{\text{b}}^{\text{high}} / A_{\text{b}}^{\text{low}} = 2.14^{+1.42}_{-0.85}$, which is $1.3\sigma$ away from unity. The lensing signal and posterior distributions for this test can be seen in Figures \ref{fig:richness_test1} and \ref{fig:richness_test2}.

\section{Discussion and conclusions}
\label{sec:conclusions}

We have measured the galaxy-galaxy lensing signal of a selection of GAMA groups split into two samples according to the radial distribution of 
their satellite galaxies. 
We use the radial distribution of the satellite galaxies as a proxy for the halo assembly time, and report 
no evidence for halo assembly bias on galaxy group scales (typical masses of $10^{13} M_{\odot}/h$). We use a halo model fit to constrain the halo masses and the large scale halo bias in order to see if the halo biases are consistent with those dictated solely by their halo masses. In this analysis, we used the KiDS data covering $180$ deg$^2$ of 
the sky \citep{Hildebrandt2016}, that fully overlaps with the three GAMA equatorial patches (G9, G12 and G15). As the 
photometric calibration and shape measurements analysis differ significantly from the previous KiDS data releases, we also perform additional tests for any possible systematic errors and biases that the new procedures might introduce {(see Appendix \ref{ref:systematics}).

Our findings are in agreement with the results from \citet{Zu2016}, who re-analysed the SDSS redMaPPer clusters sample used in \citet{Miyatake2015} and found no evidence for halo assembly bias 
as previously claimed by \citet{Miyatake2015}. They argue that that analysis suffered from misidentification of cluster members due to projection effects \citep{Zu2016}, which are minimised in the case when one uses spectroscopic information on cluster or group membership.

It is unlikely that our analysis suffers from the mis-identification of the GAMA galaxy groups members 
and/or contamination from background galaxies to the degree present in the SDSS case \citep[up to $40\%$,][]{Zu2016}, and thus artificially changing the radial distribution of the satellite galaxies. The projection effects in our case come only from peculiar velocities (and mismatching from the FoF algorithm), whereas the projection effects in \citet{Miyatake2015} are dominated by photo-$z$ uncertainties and errors, which are much larger than peculiar velocities.
If that would be the case, this would indeed have a larger effect on groups with a low number of 
member galaxies (and thus in the same regime we are using for our study).
The GAMA groups are, due to available spectroscopic redshifts, highly pure and robust -- for groups
with $N_{\text{FoF}} \ge 5$ the purity approaches $90\%$ as assessed using a mock catalogue \citep{Robotham2011}. An issue that remains is the possible fragmentation of the GAMA galaxy groups by the FoF algorithm and 
a full assessment of this potential issue is beyond the scope of this paper and we defer these topics to a study in the future.

Additionally, the assumption of a NFW profile as our fiducial dark matter density profile can potentially affect the results. Exploration of different profiles is beyond the scope of this paper, but one would not expect that the different profiles would introduce differences in the obtained halo biases. The dark matter density profile does not enter into predictions for the two-halo term which carries all the biasing information. Moreover, any systematic effects due to the differences in profile would enter into both samples in the same way, and when taking the ratio of any quantities, they would to a large extent cancel out.

In order to reach a better precision in our lensing measurements, we could use the full KiDS-450 survey area. 
This is limited however by the lack of spectroscopy to create a group catalogue. The GAMA survey will be expanded 
into a newer and upcoming spectroscopic survey named WAVES \citep{Driver2015}\footnote{Homepage: \url{http://www.wavesurvey.org}}, 
which is planned to cover the southern half of the KiDS survey ($700$ deg$^2$) and provide redshifts for up to $2$ million galaxies, which should 
provide us with enough statistical power not only to access the signatures of assembly bias in those galaxies but to extend the 
observational evidence also to galaxy scales.

\section*{Acknowledgements}

We thank the anonymous referee for their very useful comments and suggestions. A. Dvornik would like to thank to Keira J. Brooks and Christos Georgiou for proof reading the manuscript. 

K. Kuijken acknowledges support by the Alexander von Humboldt Foundation. H. Hoekstra and R. Herbonnet acknowledges support from the European Research Council under FP7 grant number 279396. R. Nakajima acknowledges support from the German Federal Ministry for Economic Affairs and Energy (BMWi) provided via DLR under project no. 50QE1103. M. Viola acknowledges support from the European Research Council under FP7 grant number 279396 and the Netherlands Organisation for Scientific Research (NWO) through grants 614.001.103. I. Fenech Conti acknowledges the use of computational facilities procured through the European Regional Development Fund, Project ERDF-080 -- A supercomputing laboratory for the University of Malta. C. Heymans acknowledges support from the European Research Council under grant number 647112. H. Hildebrandt is supported by an Emmy Noether grant (No. Hi 1495/2-1) of the Deutsche Forschungsgemeinschaft. This work is supported by the Deutsche Forschungsgemeinschaft in the framework of the TR33 `The Dark Universe'. E. van Uitert acknowledges support from an STFC Ernest Rutherford Research Grant, grant reference ST/L00285X/1. A. Choi acknowledges support from the European Research Council under the FP7 grant number 240185.

This research is based on data products from observations made with ESO Telescopes at the La Silla Paranal 
Observatory under programme IDs 177.A-3016, 177.A-3017 and 177.A-3018, and on data products produced 
by Target/OmegaCEN, INAF-OACN, INAF-OAPD and the KiDS production team, on behalf of the KiDS consortium.

GAMA is a joint European-Australasian project based around a spectroscopic campaign using the Anglo-Australian Telescope. 
The GAMA input catalogue is based on data taken from the Sloan Digital Sky Survey and the UKIRT Infrared Deep Sky Survey. 
Complementary imaging of the GAMA regions is being obtained by a number of independent survey programs including GALEX MIS, 
VST KiDS, VISTA VIKING, WISE, Herschel-ATLAS, GMRT and ASKAP providing UV to radio coverage. GAMA is funded by the 
STFC (UK), the ARC (Australia), the AAO, and the participating institutions. The GAMA website is \mbox{\url{http://www.gama-survey.org}}.

This work has made use of Python (\url{http://www.python.org}), including the packages \texttt{numpy} (\url{http://www.numpy.org}) 
and \texttt{scipy} (\url{http://www.scipy.org}). The halo model is built upon \texttt{hmf} Python package by \citet{Murray2013}. 
Plots have been produced with \texttt{matplotlib} \citep{Hunter2007} and \texttt{corner.py} \citep{cornerplot}.

\emph{Author contributions:} All authors contributed to writing and development of this paper. The authorship list reflects the 
lead authors (AD, MC, KK, MV) followed by two alphabetical groups. The first alphabetical group includes those who are key 
contributors to both the scientific analysis and the data products. The second group covers those who have made a 
significant contribution either to the data products or to the scientific analysis.




\bibliographystyle{mnras}
\bibliography{library} 




\appendix

\section{Systematics tests}
\label{ref:systematics}

We show here additional systematic tests performed as the image reduction procedure, photometric redshift calibration and shape measurement steps differ significantly from the methods used in \citet{Viola2015}. We devise a number of tests to see how the obtained data behaves in different observational limits, and the results are presented in the following paragraphs.

\subsection{Multiplicative bias}

The estimates of the average multiplicative bias $m$ for each redshift slice used in the calculation are obtained using a method presented in \citet{Conti2016}. They are further weighted 
by the weight $w^{\prime} = w_{\rm s} D(z_{\rm l},z_{\rm s}) / D(z_{\rm s})$ for a given lens-source sample. Typically, the value of the $\mu$ correction is 
around $- 0.014$, independent of the scale at which it is computed. Figure \ref{fig:m_slices} shows the estimates of the average multiplicative bias $m$ for each redshift slice used in the calculation.

\subsection{Additive bias}

Secondly, we test for the presence of the additive shear bias, by checking the tangential shear
component measured around random points. This is calculated by performing lensing measurements around $10$ million random points 
in RA and DEC (for all three GAMA patches), which have the same assigned redshift distribution as the GAMA galaxies. We use 
version 1 of the GAMA random catalogue, created as described in \citet{Farrow2015}. 
Like the cross component of the measured ellipticities, also the azimuthally averaged tangential shear signal around random points should equal to zero. 
Figures \ref{fig:randoms} and \ref{fig:cross} show significant systematic errors on scales larger than $1$ Mpc/$h$ as well as patch-dependent systematic errors. 
We perform the analysis on three patches separately (G9, G12 and G15). As discussed in \citet{Hildebrandt2016} 
and \citet{Conti2016}, the correction for the additive bias obtained using image simulations should only be obtained for 
individual KiDS patches, due to specific systematics associated with each patch.  We also check for the behaviour of the cross shear component. 
Any presence of the cross component signal points towards the presence of systematic errors and thus measurements on scales with significant cross component signal have to be corrected before using them for scientific purposes.

One could estimate the additive bias using image simulations \citep[using a method shown in][]{Conti2016}, but that will only account for the PSF effects.
We correct for the additive bias using the results 
obtained from the random signal as the additive bias might arise because of spurious objects (including asteroids, stellar spikes, pixel defects, etc.) in our lensing data, apart from PSF effects. It is thus important to correct for it using 
the data. 
Correction of additive bias is performed by subtracting the random signal obtained for each patch from the true ESD measurement in the same patches. Doing so, that also gives better covariance matrix estimates \citep{Singh2016}.
The final ESD profile is calculated by combining the random-subtracted signals from all three patches.

\begin{figure*}
	\centering
	\includegraphics[width=\textwidth]{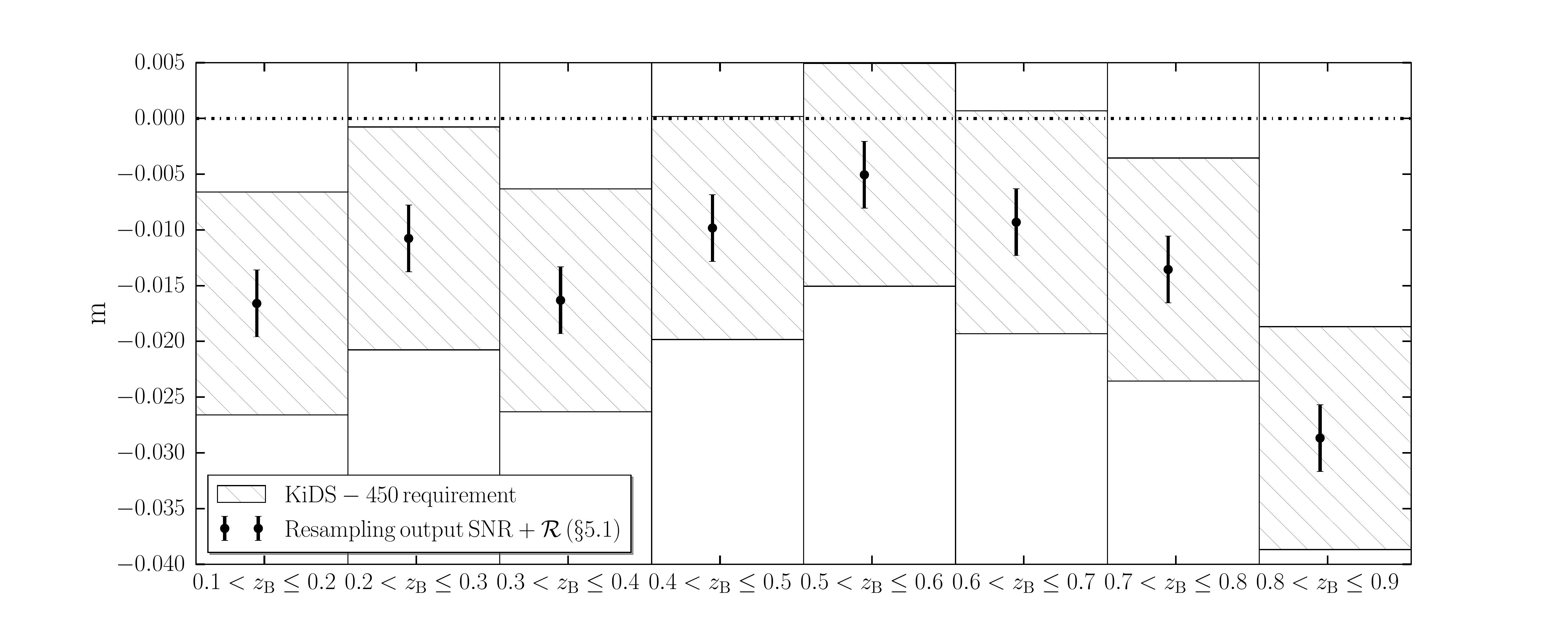}
 	\caption{Multiplicative bias calculated using the resampling technique of \citet[][chapter $5.1$]{Conti2016} in the redshift 
	slices used in this analysis. The hatched area indicates the requirement on the knowledge of the multiplicative bias for KiDS-450 cosmic shear analysis \citep{Hildebrandt2016}.}
	\label{fig:m_slices}
\end{figure*}

\begin{figure}
	\centering
	\includegraphics[width=\columnwidth]{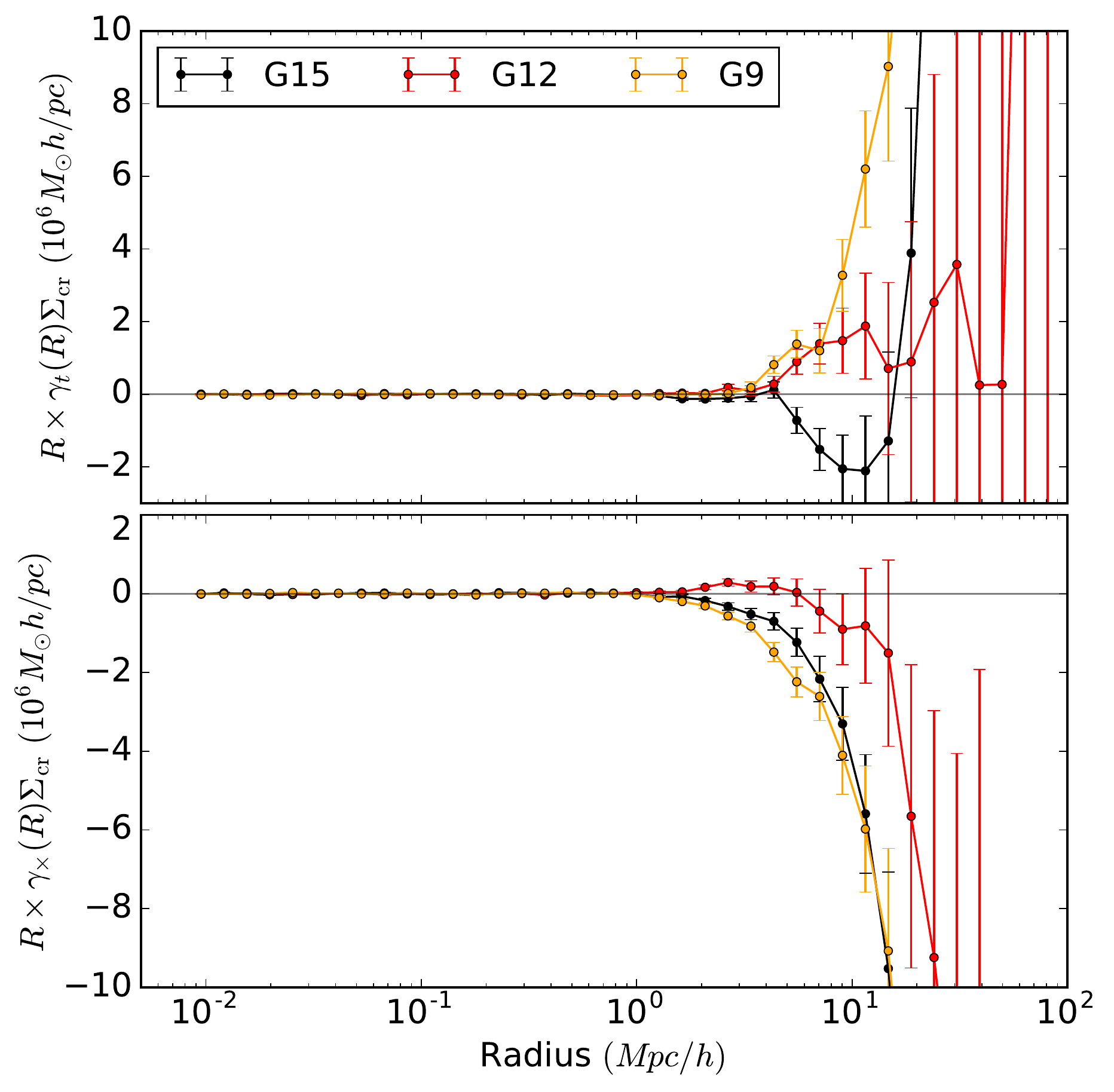}
 	\caption{Shear signal around 10 million random points having the same redshift distribution as GAMA galaxies, 
	split between the three GAMA patches. Shown are both tangential ($\gamma_{t}$, upper panel) and 
	cross ($\gamma_{\times}$ lower panel) components. We use these 
	measurements to correct for the additive bias in our measured ESD signal.}
	\label{fig:randoms}
\end{figure}

\begin{figure}
	\centering
	\includegraphics[width=\columnwidth]{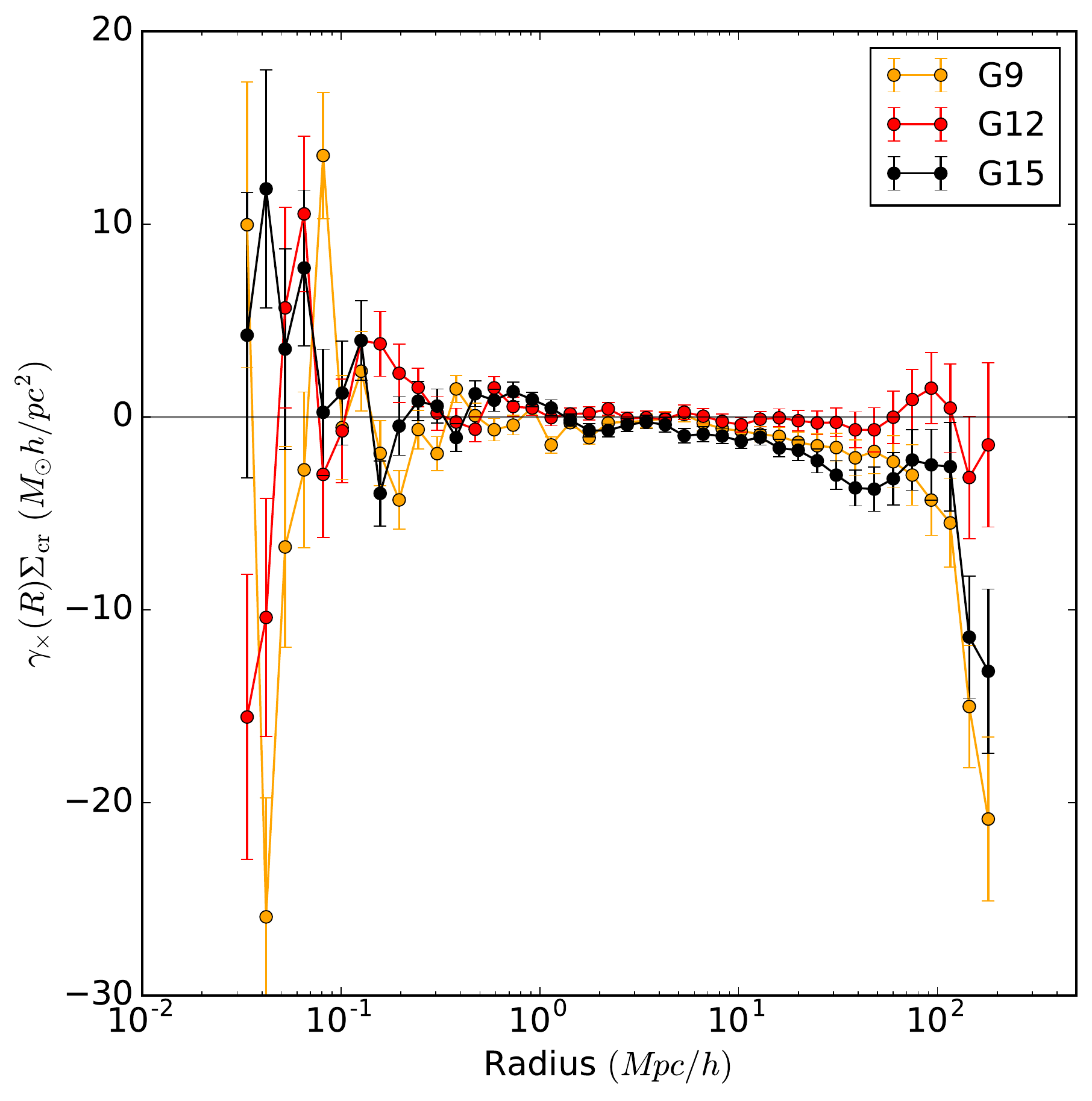}
 	\caption{Lensing signal computed from the cross component of measured ellipticities, around all GAMA galaxies in the 
	three equatorial patches (G9, G12 and G15). One can see, that the systematic errors significantly affect the signal 
	below $70$ kpc/$h$ and above $10$ Mpc/$h$, with the G12 patch being the least affected, even after subtracting the signal computed around random points.}
	\label{fig:cross}
\end{figure}

\subsection{Group member contamination of the source galaxies}

The next important test we perform is to check how much the GAMA galaxy group members contaminate our source population 
\citep[the so-called \emph{boost factor};][]{Miyatake2013, VanUitert2016a}. Those galaxies will dilute the lensing signal (as they are not lensed).  
The resulting lensing signal will be biased (Figure \ref{fig:overdensity}) on small scales with the source over-density up to 
$30\%$ at $75$ kpc/$h$ (Figure \ref{fig:overdensity}). We can impose a more stringent cut 
than the cut $z_{\rm s} > z_{\rm l}$ used in previous studies on KiDS and GAMA data, by adding an offset $\delta_{z}$ to the cut on the source population. As seen in Figure \ref{fig:overdensity}, 
using a conservative cut with $\delta_{z} = 0.1$ still leaves a $10\%$ over-density in the source sample. More conservative cuts 
lower the observed over-density, as expected. They also suppress the contamination, but this is not ideal as real source galaxies are removed as well, since it decreases the lensing signal-to-noise.On the small scales (below $75$ kpc/$h$) the decrease of the source density is connected with the fact that the source galaxies become obscured by the host BCG of the GAMA group. The ESD signals in Figure \ref{fig:overdensity} are corrected with the boost factor using the factors shown in the top panel of the same figure and have lensing efficiency calculated separately for each redshift cut. We find that for a redshift offset of $\delta_{z} = 0.2$ the boost correction is not necessary.

\begin{figure}
	\centering
	\includegraphics[width=\columnwidth]{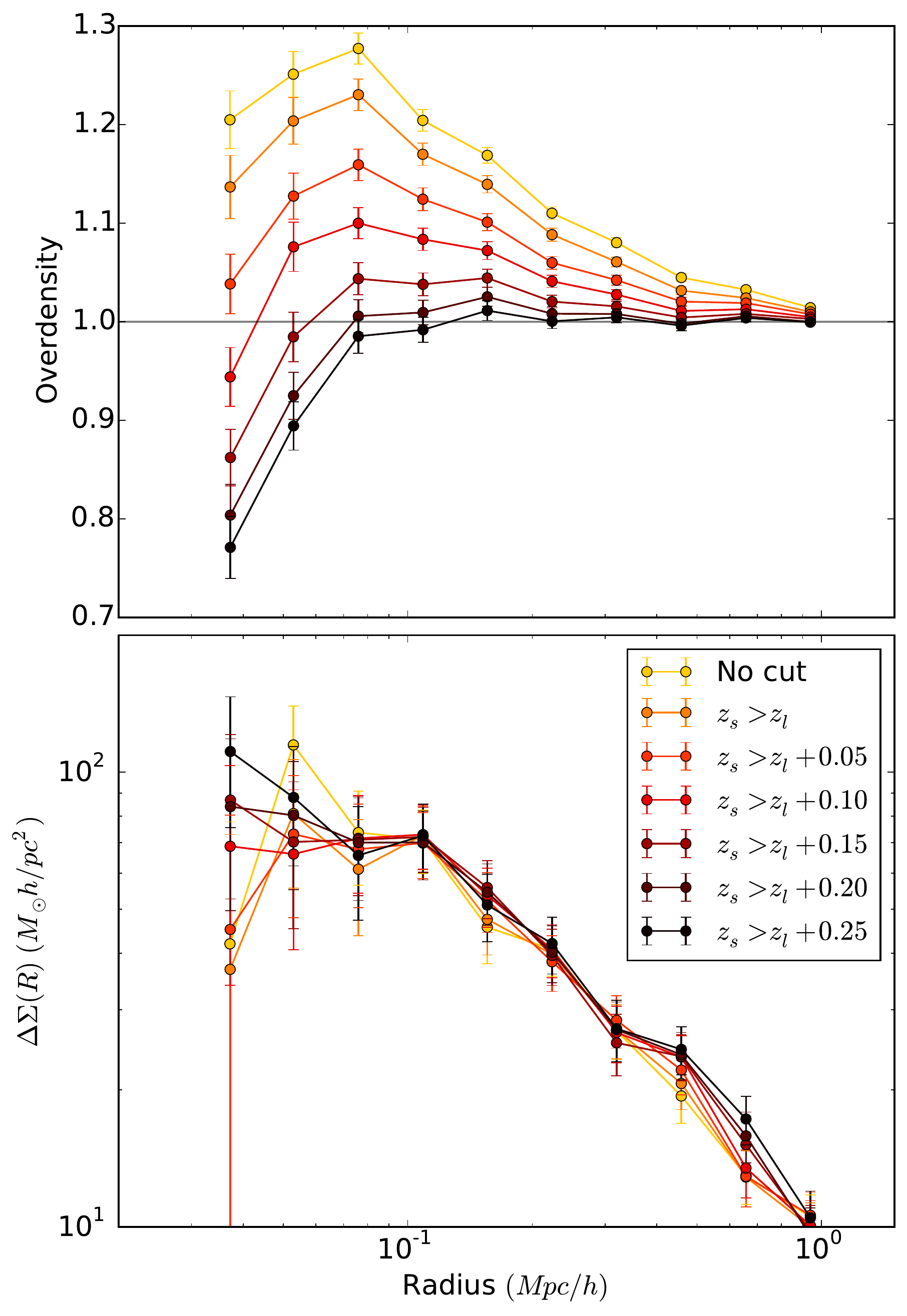}
 	\caption{\textit{Top panel:} The overdensity of KiDS source galaxies around GAMA galaxy groups with richness $N_{\text{FoF}} \ge 5$. 
	The various lines correspond to different redshift cuts applied to the source sample. Even for a conservative cut of 
	$z_{\rm s} > z_{\rm l} + 0.1$, we find a residual contamination of group members in the source sample of up to $10\%$ at $75$ kpc. 
	\textit{Bottom panel:} The ESD signal around GAMA galaxy groups with richness $N_{\text{FoF}} \ge 5$ up to $2$ Mpc/$h$. The various 
	lines correspond to different redshift cuts applied to the source sample. The redshift cut does not significantly affect the lensing signal, 
	but one removes any possible problems due to group contamination. The lensing signals are computed using different lensing efficiencies and are corrected 
	with the boost factor using the factors shown in the top panel.}
	\label{fig:overdensity}
\end{figure}

\subsection{Source redshift distribution}

The significant difference between this analysis and previous 
method presented in \citet{Viola2015} is the usage of full redshift probability 
distribution of the sources, $n(z_{\rm s})$, compared to \citet{Viola2015} where each source is given its own posterior redshift distribution $p(z_{\rm s})$ obtained from BPZ. With the following tests we want to see what the difference between having only the global $n(z_s)$ has on the error budget and the resulting lensing signals.
The observable lensing signal depends on the angular diameter distances to the lens and source galaxies (Equation \ref{eq:crit_effective}). 
The redshifts to the lens galaxies are known from the GAMA spectroscopic survey, while for the sources 
we need to resort to the photometric redshifts derived using multi-band images (in $ugri$ photometric bands) of the KiDS survey. 
The colors obtained using those images are a basis for the photometric redshift estimates, which also provides us the full redshift probability 
distribution of the sources, $n(z_{\rm s})$, obtained using the direct calibration method 
\citep[for more information and comparison with other techniques see][]{Hildebrandt2016}. 
Comparison between the final lensing signals using the individual $p(z_{\rm s})$, the stack of $p(z_{\rm s})$ and the global $n(z_{\rm s})$ can be seen in the bottom panel of Figure \ref{fig:nz_pz_esd} and
the difference between the stacked $p(z_{\rm s})$ and $n(z_{\rm s})$ probability distributions in the top panel of the same Figure. 
The resulting lensing signals do not change much, and are all in agreement within the error budget of the lensing signal of all the GAMA galaxies. 
Following \citet{Hildebrandt2016}, we adopt the redshift range $[0.1, 0.9]$, which is the same as the 
covered range by the $4$ tomographic bins used in \citet{Hildebrandt2016}. 

\begin{figure}
	\centering
	\includegraphics[width=\columnwidth]{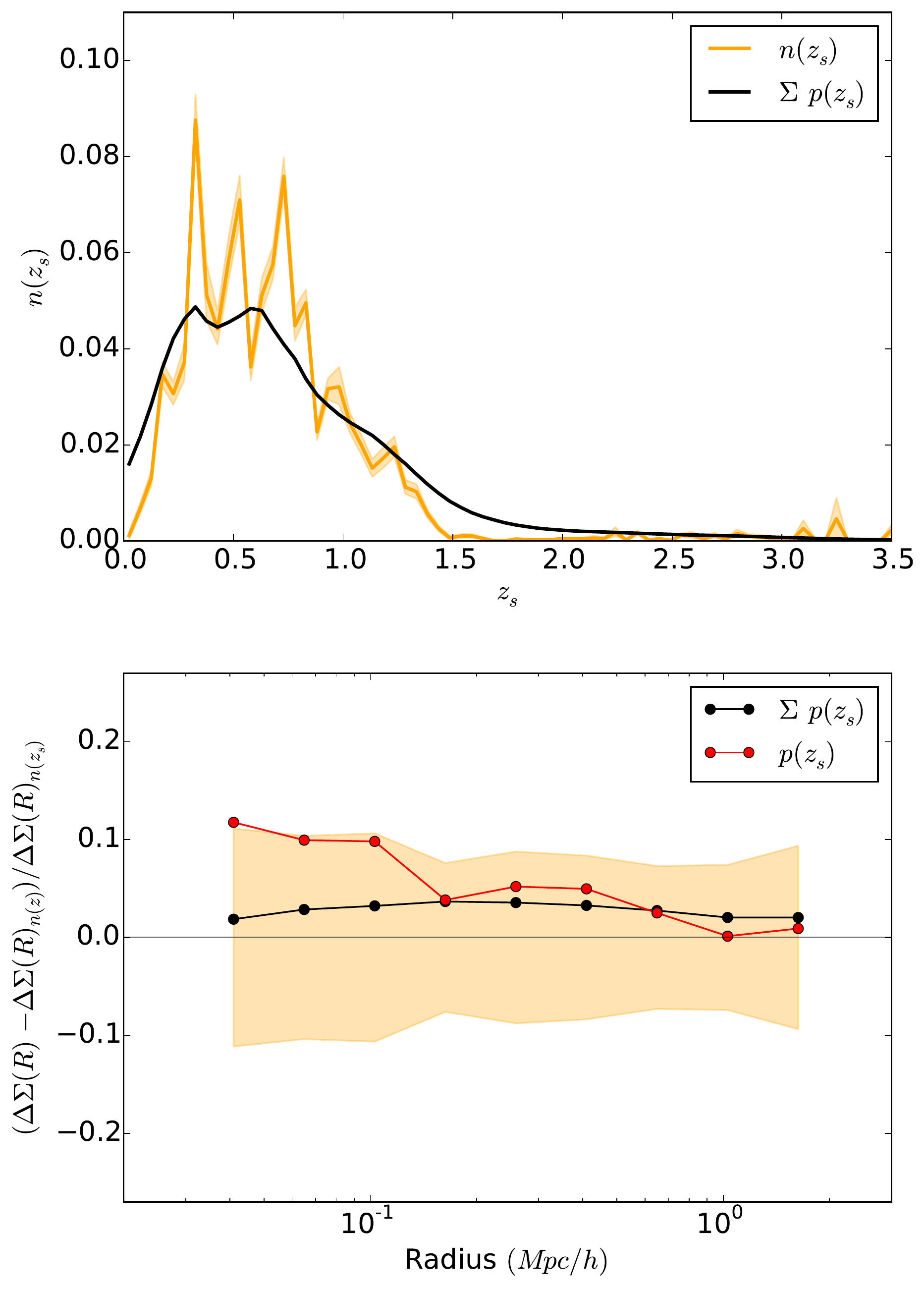}
 	\caption{\textit{Top panel:} Comparison of the $n(z_{\rm s})$ as given by the direct calibration method (DIR) and the stacked $p(z_{\rm s})$ 
	obtained from BPZ \citep{Hildebrandt2016}. As already noted in \citet{Hildebrandt2016}, the stacked $p(z_{\rm s})$ does not accurately 
	reproduce the features seen in the DIR method, and its usage is discouraged. \textit{Bottom panel:} Difference between 
	the lensing signal using three different source redshift distributions. $p(z_{\rm s})$ represents the method as used in \citet{Viola2015}, 
	compared to the stacked $p(z_{\rm s})$ and the $n(z_{\rm s})$ obtained using DIR (for all $180,960$ GAMA galaxies). 
	Within the error budget, all the methods are in agreement (the orange area is the error on the lensing signal calculated using the $n(z_{\rm s})$).}
	\label{fig:nz_pz_esd}
\end{figure}

The uncertainty on the $n(z_{\rm s})$ contributes to the total error budget of the lensing signal. As the errors due to this uncertainty 
can affect the conclusions of the quantitative results, we look into how much the actual contribution is. We take $1000$ bootstrap 
realisations of the weighted spectroscopic catalogue \citep{Hildebrandt2016} giving us $1000$ different realisations of $n(z_{\rm s})$, 
for which we calculate the lensing signal. This gives us enough samples to constrain the uncertainty on the lensing signal due to 
the uncertainty on the $n(z_{\rm s})$. We compare the given $1\sigma$ errors with the total error on our lensing signal. 
The results can be seen in Figure \ref{fig:sn_z_tot}, where it is clearly seen that the uncertainty on $n(z_{\rm s})$ is 
sub-dominant to the whole error budget.

\begin{figure}
	\centering
	\includegraphics[width=\columnwidth]{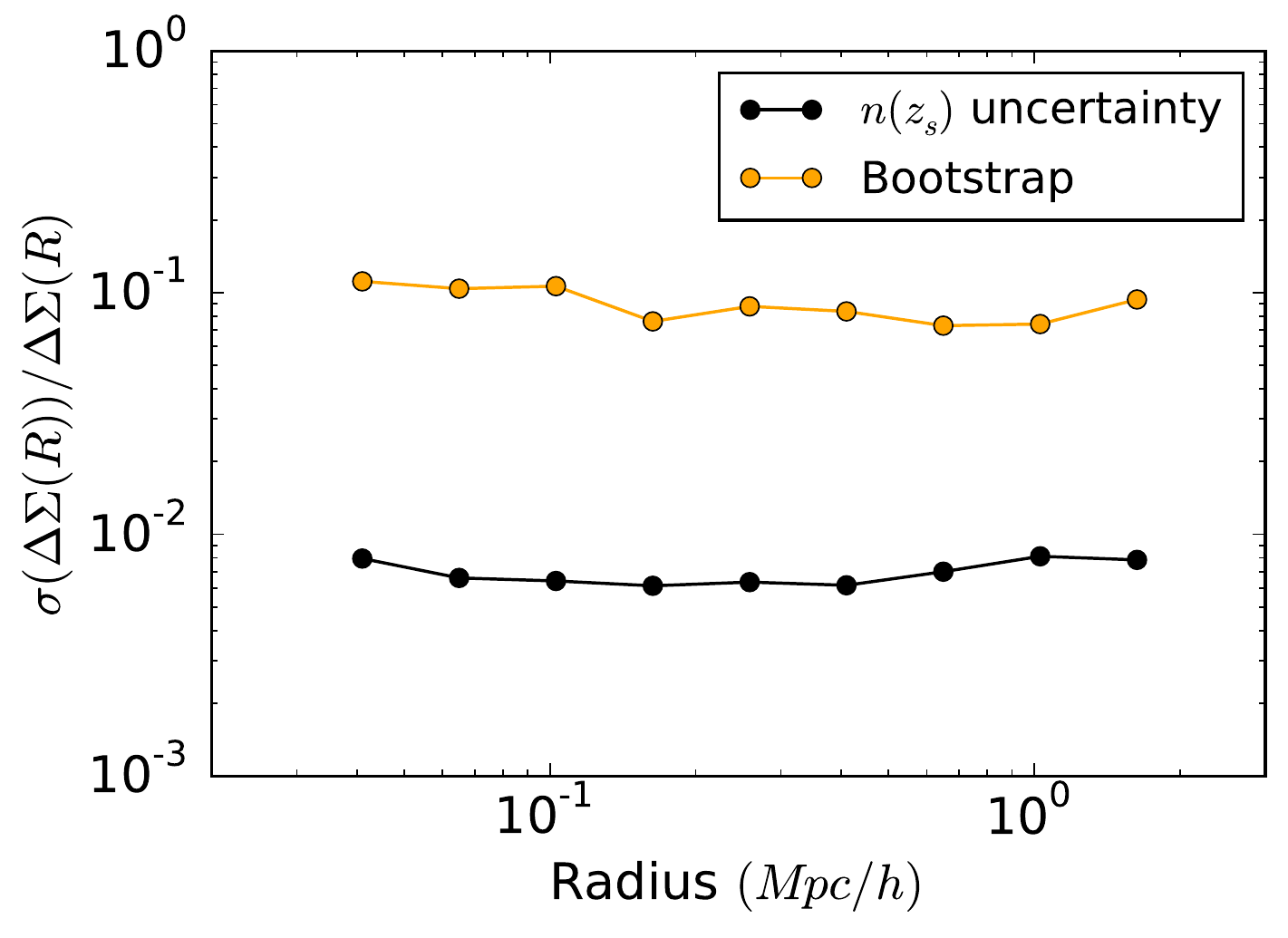}
 	\caption{Relative error estimates of the $n(z_{s})$ uncertainty compared to the uncertainty as obtained using the bootstrap method 
	on the lensing signal (including shape noise and cosmic variance contributions), calculated for the full sample of GAMA galaxies 
	in the three equatorial patches (G9, G12 and G15). It can be seen that the contribution to the total error 
	budget from the uncertainty of the redshift distribution is negligible.}
	\label{fig:sn_z_tot}
\end{figure}

\section{Full posterior distributions}
\label{sec:more_plots}

Figures \ref{fig:richness_test1} and \ref{fig:richness_test2} show lensing signal and posterior distributions of the additional test of splitting the full sample to two samples with high and low richnesses (as discussed in Section \ref{sec:res_lensing}). In Figure \ref{fig:triangle_full} we show the full posterior probability distribution for all fitted parameters in our 
MCMC fit as discussed in Sections \ref{sec:halomodel} and \ref{sec:res_lensing}.

\begin{figure}
	\centering
	\includegraphics[width=\columnwidth]{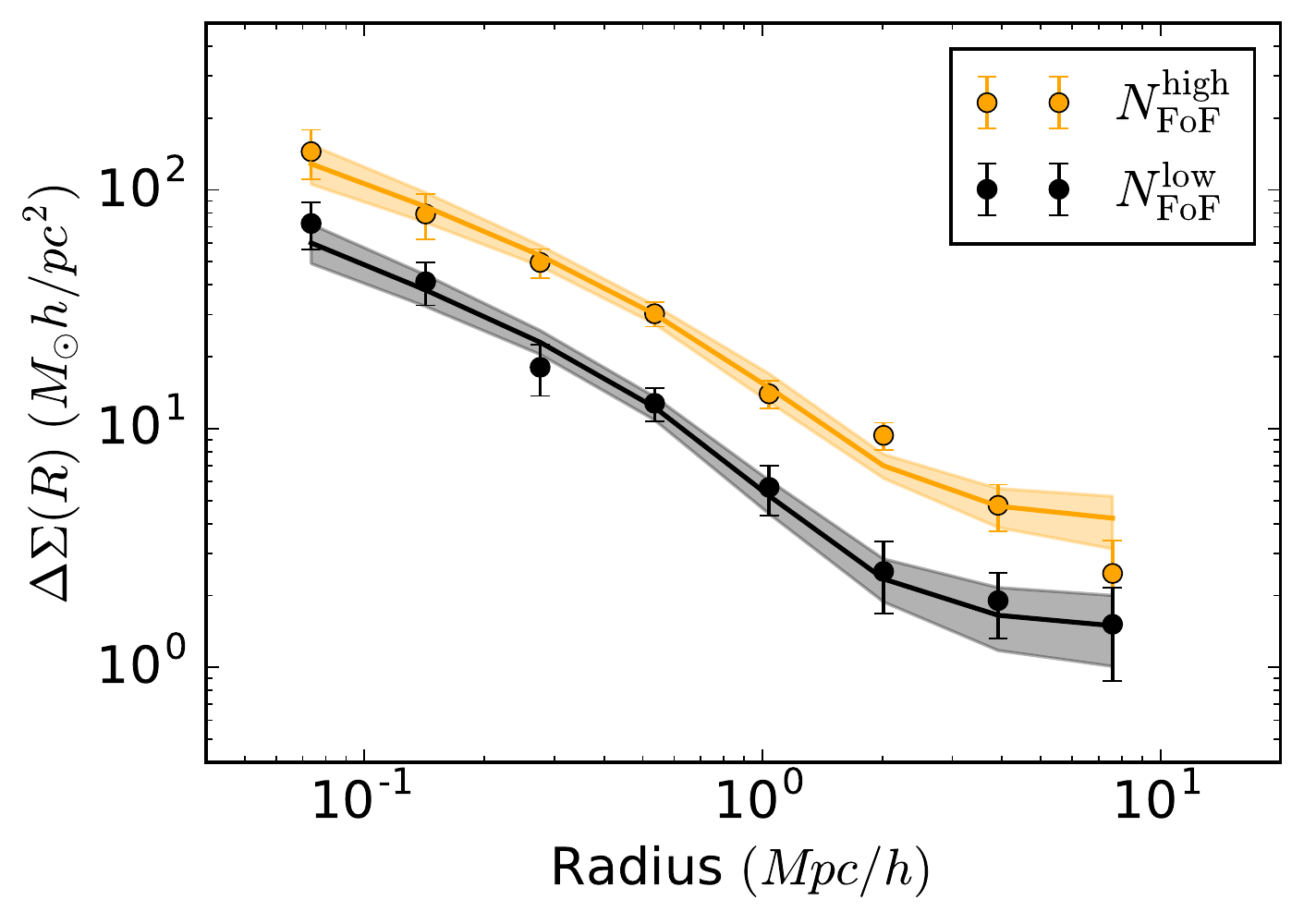}
 	\caption{Stacked ESD profiles measured around the central galaxies of GAMA groups, selected according to the apparent richness of the groups. 
	The solid orange and black lines represent 
	the best-fitting halo model as obtained using a MCMC fit, with the $68 \%$ confidence interval indicated with a shaded region.}
	\label{fig:richness_test1}
\end{figure}

\begin{figure}
	\centering
	\includegraphics[width=\columnwidth]{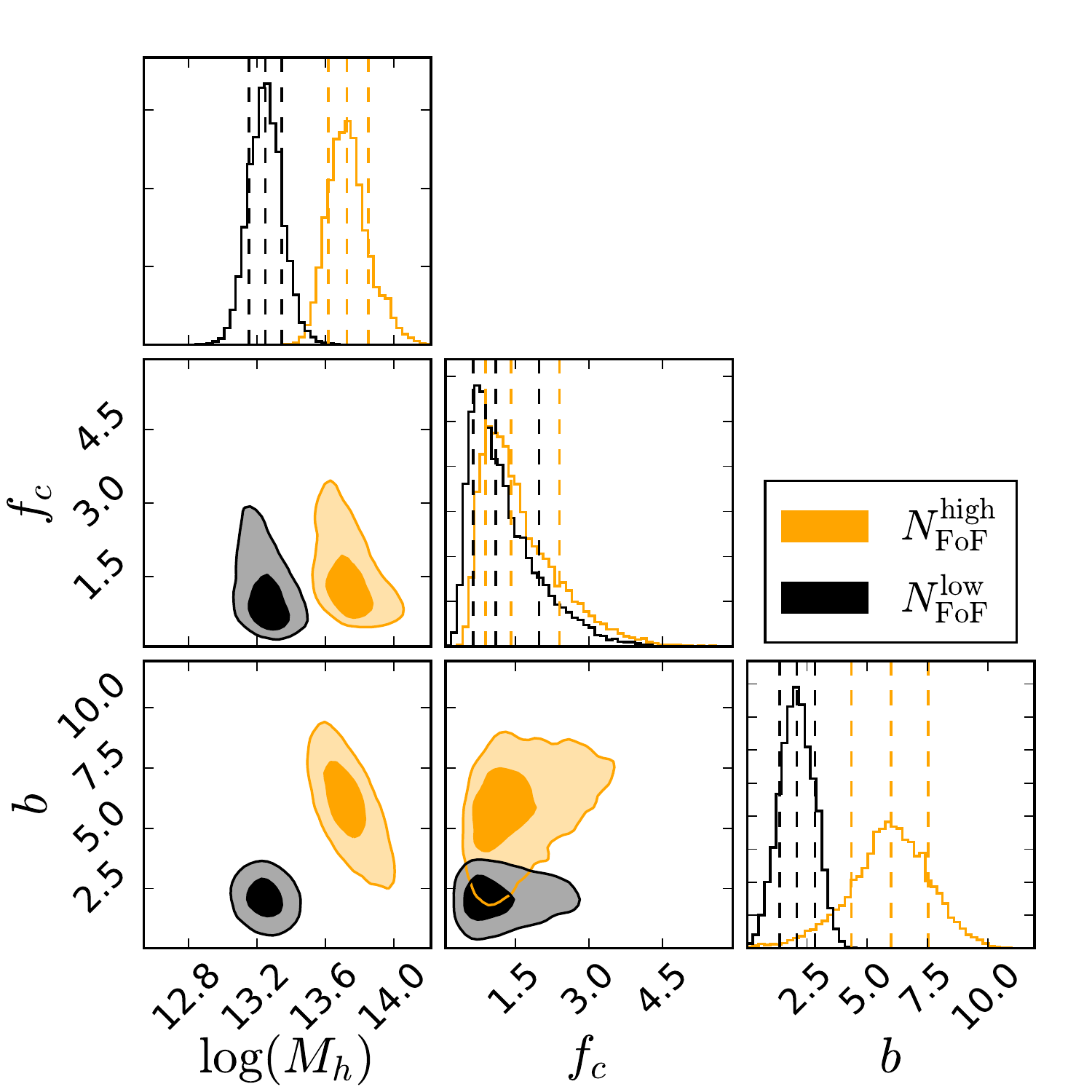}
 	\caption{The posterior distributions of the halo model parameters $M_{h}$, $f_{c}$ and $b$ for the sample of lenses split 
	according to their apparent richness. 
	The posterior distribution clearly shows a difference in the obtained halo masses as well as a significant 
	difference in the obtained halo biases. The contours indicate $1\sigma$ and $2\sigma$ confidence regions.}
	\label{fig:richness_test2}
\end{figure}

\begin{figure*}
	\includegraphics[width=17cm]{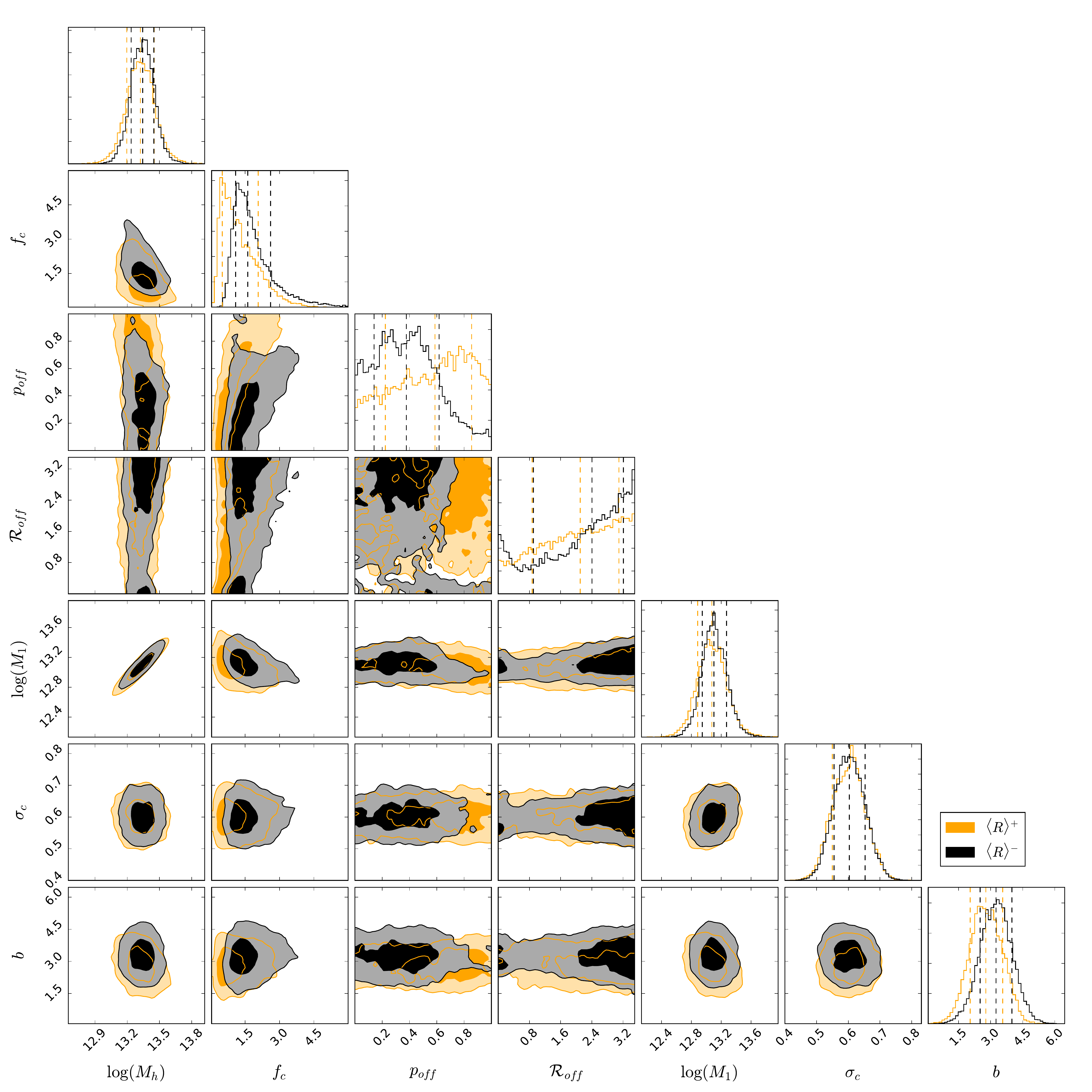}
 	\caption{The full posterior distributions of the halo model parameters $M_{h}$, $f_{c}$ $p_{\text{off}}$, ${\cal R}_{\rm off}$, $M_{1}$, $\sigma_{c}$ and $b$. 
	The posterior distribution clearly shows a slight difference in the obtained halo masses as well as no 
	difference in the obtained halo biases, the miscentering parameters and the normalisation of concentration-halo mass relation. 
	The contours indicate $1\sigma$ and $2\sigma$ confidence regions. 
	Priors used in the MCMC fit can be found in Section \ref{sec:halomodel}.}
	\label{fig:triangle_full}
\end{figure*}


\bsp	
\label{lastpage}
\end{document}